\documentstyle[aps,epsf]{revtex}

\begin{document} \renewcommand{\thefootnote}{\fnsymbol{footnote}}

\draft \twocolumn[\hsize\textwidth\columnwidth\hsize\csname@twocolumnfalse\endcsname]
\title{Model of macroeconomic evolution in stable regionally dependent economic fields}

\author{M. Ausloos$^1$, P. Clippe$^2$ and A. Pekalski$^3$ }

\address{ $^1$ GRASP and SUPRATECS, B5, Sart Tilman,  B-4000 Li\`ege, Belgium \\$^2$ GRASP, B5, Universit\'{e} de Li\`{e}ge,  B-4000 Li\`{e}ge, Belgium, \\$^3$ Institute of Theoretical Physics, University of Wroclaw, pl. Maxa Borna 9,
PL-50-204 Wroclaw, Poland }

\date{\today} 
\maketitle

\begin{abstract}

We develop a model for the evolution of economic entities within a geographical
type of framework. On a square symmetry lattice made of three (economic) 
regions, firms, described by a scalar fitness, are allowed to move, adapt, 
merge or create spin-offs under predetermined rules, in a space and time 
dependent economic environment.  We only consider here  one timely variation 
of the ``external economic field condition''. For the firm fitness evolution 
we take into account a constraint such that the disappearance of a firm 
modifies the fitness of nearest neighboring ones, as in Bak-Sneppen population 
fitness evolution model. The concentration of firms, the averaged fitness, the 
regional distribution of firms, and fitness  for different time moments, the 
number of collapsed, merged and new firms as a function of time have been 
recorded and are discussed.  Also the asymptotic values of the number of 
firms present in the three regions together with their average fitness, as 
well as the number of respective births and collapses in the three regions 
are examined. It appears that a sort of $critical$ selection pressure exists.  
A power law dependence, signature of self-critical organization is seen in 
the birth and collapse asymptotic values for a high selection pressure only. 
A lack of self-organization is also seen at region borders. 
\end{abstract}
 

\footnotetext[1]{electronic address: marcel.ausloos@ulg.ac.be}
\footnotetext[2]{electronic address: P.Clippe\@ulg.ac.be}
\footnotetext[3]{electronic address: apekal\@ift.uni.wroc.pl}

\section{INTRODUCTION}

The present contribution extends our earlier paper,\cite{ACP,ACP04}. World
economic conditions evolved and are quite varied on different time and space
scales. Basic questions  are whether the consequences of political conditions
have predictable effects, or not and whether annoying situations can be avoided.
The questions pertain to macroeconomic themes, not so often touched upon in
econophysics  (see exceptions in\cite{Richards,Gligor,Aoki2}).

The case of the Berlin wall fall followed by Eastern Europe and Central Asia
market openings to a so called liberal economy is a intriguing event. From an
econophysicist point of view, the event(s) can be considered as an increase in
"physical volume, or available space" as well as in a modification of the {\it
external fields}\cite{ACP}. In the latter reference we have discussed whether one
can describe, within a simple model, the concentration of enterprises and their
so called "fitness" as a function of time and space, under varying in time and
space economic field conditions.  We have observed a non-trivial behavior with
cycle features; in our opinion more pronounced that those discussed by Aoki,
Kalecki, Freeman and others\cite{Aoki2,AokiShirai,Kalecki,Freeman}. The model
however seemed to be too optimistic  in not allowing enough bankruptcies,
collapses or mere disappearance of firms in the process, in particular with
respect to the so called economic selection pressure\cite{ACP,ACP04}. Moreover we
neglected the fact that a firm disappearance could locally modify the probability
of survival of neighboring ones. Indeed an immense political worry concerns the
effect of degradation of an economy through avalanche-like processes.

In practice, adaptation to a dynamic  environment contains a trade-off: it is
more difficult to adapt in  the next (time) step as shown e.g. also on financial
time series by Yamasaki et al.\cite{Kazuko}. It is also expected that the
adaptive behaviors of  ''populations''  differ depending whether a    Darwinian
or Lamarckian scheme is implied. A large discussion exists whether such schemes
hold or not, and how in economy, e.g. see
\cite{Alchian,Hirshleifer,Boulding,Eldredge3,Kelm,Hogdson,BerghGowdy01,NelsonWinter82}
for references and a recent review.  Alas, the definition of Lamarckism (or
Darwinism) is not totally agreed upon in economy circles.

Therefore we present and discuss here below a more elaborate model than in
Ausloos {\it et al.}\cite{ACP,ACP04}, i.e. in order to describe changing basic
economic conditions and implement more realistic Darwinian-like adaptation rules,
as discussed in economy
circles\cite{Alchian,Hirshleifer,Boulding,Eldredge3,Kelm,Hogdson,BerghGowdy01,NelsonWinter82}
- in space and time. In essence we introduce a Darwin-Bak-Sneppen-like
constraint\cite{BakSneppen} in our previous macro-economy\cite{ACP,ACP04}
microscopic physics-like model, i.e. the choice of the initial firm or company
depends on its relative fitness with respect to the external field as
in\cite{ACP04};  the chosen firm is the less adapted one, instead of being chosen
at random\cite{ACP}.  The fitness of neighboring ones evolve according to some
{\it a priori} defined dynamics.

The notion of economic external field is also important. Nelson and
Winter\cite{NelsonWinter82} write that ''at any time, firms in an industry can be
viewed as operating with a set of techniques and decision rules (routines) keyed
to conditions external to the firm'' ... ''and to various internal state
conditions''. Surely self-organization and external (field) conditions exist at
the same time, and are  hard to separate\cite{Foster97,Silverberg}. However there
are also many different conditions affecting many characteristic variables of a
company or industry state.  Thus searching for as simple as possible description
and consequences of such external though often qualitatively only apparent
constraint, to be contrasted to self-organization is of interest. As usual in
physics  when intending extreme simplification this field should be applied
toward or coupled to some variable such that some function be
optimized\cite{Radner}, like the free energy in thermodynamics.  In macro and
micro-economy the field can be understood as of political origin, but also of
more physical origin like the weather. In the latter case, see the weather
influence on the energy prices\cite{Weron}. As example of the former one can
think about deregulation, or wage and benefit taxation influences upon
localization and production or market approach by a company.

Along this line, the most extreme simplification leads to represent a company by
a variable having the same number of characteristic symmetry group elements as
the field. Thus a scalar, called the fitness of a company is
introduced\cite{ACP,ACP04}. One can consider that the fitness is the price of
stocks and the external field can be in some sense the value of the S\&P500 to
which the price is compared. A $buy/sell$ strategy will depend on the difference
between the two values. A too large difference can suggest bankruptcy and
disappearance of a company, or  the 'need'' for a modification of the business
plan. The notion of fitness of a company is related to such a book value. The
fitness might also represent the number found on gross sales or benefit lines....
The field notion has to be subsequently adapted.

Any such system will be evolutionary if it has a fourth ingredients: beside the
set of degrees of freedom, each ranked by a quality fitness criterion, and a
mechanism for introducing mutations to interject diversity into the  productive
process, the system must be governed by a selection process based on the fitness
ranking\cite{AokiShirai,Aoki1,JMMALadek} as in biology\cite{APKSV}.

Thus we consider a system composed of firms located at the nodes of a locally
square symmetry lattice of  $L_x, L_y$ size dimensions. As in Ausloos {\it et
al.}\cite{ACP} the system is divided into three parts, $k$= I, II and III,  by
vertical boundaries at $L_{xk}$, i.e. lines along the $y$ axis at
$L_{x1},L_{x2}$. There is no periodic boundary conditions. At the beginning there
are $N(t=0)$ firms, located only in the first (I) region. One lattice site may be
occupied by only one firm. Apart from its location on the lattice a firm $i$ is
characterized by its scalar fitness, $f_i$, taken from a uniform distribution,
$\in $ (0,1). There is no other factor distinguishing the firms, hence we may
consider them to belong to only one group of industrial companies, i.e.
representing a particular industrial branch. Initially the fitness values are
attributed to firms in a random way.

Any firm may change  its position\cite{Silverberg} to a vacant nearest
neighboring site chosen at random and may either merge with another firm (with a
given probability, - according to its business plan) or create a new one. In a
region, a firm is under an {\it external field} $F$ $ \in$ (0,1), which
characterizes the local economic conditions. The local  political system may be
more or less demanding,- more or less high taxes, salary rules, worker age (and
other) conditions, ...;  this is taken into account through the parameter, $sel$,
- corresponding to selection in biological systems \cite{APKSV}. In our model it
is also a scaling parameter for the field and the fitness, favoring or not the
possible disappearance of a company unfit to its environment, the
field\cite{Kauffman2000,Auerswald2000,Frenken}.

The field $F$ may (and will)  differ from region to region and may also change in
time, measured in Monte Carlo Steps (MCS). In the following we will only consider
one drastic field change and keep stable (or static) all environmental (economic
field) conditions for evaluating its effect on the population of entities
evolution.

As in Ausloos {\it et al.}\cite{ACP} after some time the barrier separating the
region I from regions II and III is open and the diffusing firms may enter the
latter regions. At that time the external field also changes in the first region
and some values are given to the fields in the regions II and III.

Before any activity is taken for the firm (moving, merging...) its condition is
checked by comparing its fitness with the field $F$. If the two values differ
strongly, the firm may collapse and be removed from the system (see the algorithm
below).

The Bak-Sneppen condition, i.e. a modification of the  fitness of the
$eliminated$ firm nearest neighbors, is implemented through distributing new
(uncorrelated in space and time) random fitness to these   (at most 4) neighbors.
This corresponds to e.g. simulating the effect a bankruptcy may have on other
firms connected to the one which went down. Obviously the fitness neighbors can
increase or decrease. Some firms may profit or regret the disappearance of a
competitor in the local market.

Firms will  also be systematically moving on the lattice, looking for partners in
view of merging or producing spin-offs.  We will take into account the number of
contacts in the company network (neighbors on a lattice in our case) for
describing the firm evolution. It is empirically known that the replacement of
some  contacts is needed for
innovation\cite{Kauffman2000,Auerswald2000,Frenken,Aoki3}.

The (new) firm(s) fitness evolution rule is slightly different from that in
Ausloos {\it et al.}\cite{ACP}, but still takes into account the fitness of
parents.

The following quantities are recorded: concentration of the firms in the three
regions and the averaged fitness in the three regions both as a function of time,
the spatial distribution of firm concentration and fitness for different time
moments, the number of collapsed, merged and new firms as a function of time,
finally the asymptotic, i.e. when a stationary state is reached, values of the
number of firms present in the three regions together with their average fitness,
as well as the number of respective birth and collapses.

In section 2, we outline the simulation technique used for implementing the
model. We present a few results in Sect. 3, and end with a conclusion in Sect.4.

\section{MODEL}

The algorithm we implemented is the following one:

\begin{enumerate}

\item from a number of firms $N(t)$ at a given time $t$ a firm $j$ is randomly
picked

\item its survival probability is calculated from the condition

\begin{equation} p_j = \exp\left(- sel |F - f_j|\right) \end{equation}

and compared to  a random number $r$ taken from a uniform distribution $\in $
(0,1). If $r > p_j$,  the firm collapses, i.e. it is removed from the system; the
nearest neighbors of the eliminated firm receive new $random$ numbers (taken from
a uniform distribution) for their fitness. The algorithm then goes back to 1,

\item  if $r < p_j$, we try to move the firm to a nearest neighborhood (NN). A
random number, $r_1$, is generated from a uniform distribution and if it is
smaller than 0.25 then we check whether  the ''Northern NN'' is an empty site,
and we move the firm there, if possible; if $r_1$ is between 0.25 and 0.50 then
we look into the Western NN, etc. If the displacement trial is not successful the
algorithm goes back to 1 and search for a new firm;

\item next we look for a partner in the nearest neighborhood of the displaced
firm new position. If there is a firm  in the NN,

\item with a probability ($b$ =) 0.01 the neighbor, say $i$, merges with the old
firm $j$, which changes its fitness to

\begin{equation} f_j = 0.5 (f_i + f_j + (0.5 - r_2) |f_i - f_j|),\end{equation}
where $r_2$ is a random number in the range (0,1). The firm $i$ then disappears
from the system;

\item otherwise with a probability ($1-b$ =) 0.99  the firms $i$ and $j$ produce
a new firm $k$, a $spin-off$. The $k$ firm is randomly positioned in the Moore
neighborhood of the $j$ firm on an empty site if any exists. The procedure for
finding an empty place is similar to the one used when looking for an empty site
to move the firm, except that on a square lattice the Moore neighborhood consists
of 8 sites - NW, N, NE, W, E, SW, S and SE; whence if the random number $r_3$ is
smaller than 0.125 we check the site NW, if it is larger than 0.125 but smaller
than 0.250 we check the N site, etc. The new firm receives its fitness depending
on that of both parents (which remain in the system) as

\begin{equation} f_k = 0.5 (f_i + f_j + (0.5 - r_4) |f_i - f_j|),\end{equation}
where $r_3$ and $r_4$ are random numbers in the range (0,1).

\item When $N(t)$ firms were picked through (1), one MCS is said completed.

\end{enumerate}

Typical values taken for the discussed simulations below were : $L_{x1}$ = 50,
$L_{x2}$ = 100, $L_y$ = 201, $c(0)$ $\simeq$ 0.8, corresponding to a number of
firms $N(0)$  = 8040.

The parameters of the model are: $sel$, selection pressure; $t_{change}$, time
after which the barrier at $L_{x1}$ is open and the field changed in I; the
values of the fields. At the beginning the field in region I has the value $F_I$
= 0.5, which is ''optimal''; the values in the regions II and III are irrelevant
before the opening of the $L_{x1}$ barrier, since there are no firms there. After
the change, at time $t_{change} $ = 100 MCS, the values were arbitrarily taken to
be: $F_I$ = 0.3, $F_{II}$ = 0.5, $F_{III}$ = 0.6, which means that the conditions
in the region I deteriorated and the best situation is in region II. The
conditions in region III are also imposed to be better than those in region I.
These values remain unchanged till the end of the simulations.

Although the presented curves were obtained each from a single simulation, we
have checked that a different initial distribution of the firms (for the same
initial concentration, external field and selection) leads to a very similar,
even quantitatively, set of situations. We are not interested here in the
transient regime before the first 100 MCS. The simulations were carried out till
a stationary state was reached, when each investigated quantity mildly oscillated
around a rather stable average value. This happened quite soon (several hundreds
of MCS) for low selection, but we had to run till 20000 MCS for the highest
selection $sel$ = 1.7. For a $sel$ as ''high '' as  1.8, we found that no firm
survived after the initial 800 MCS in the whole system.

\section{RESULTS and COMMENTS}

There are several ways of presenting pertinent results.  In the following we
stress cases demonstrating the pertinence of the model in view of economic
qualitative observations. The scales are chosen  in order to allow for better
visually comparing cases and emphasizing differences in behavior.

Recall that from time $t_{change} $ = 100 MCS, the field values are   kept to be
$F_I$ = 0.3, $F_{II}$ = 0.5, and $F_{III}$ = 0.6 respectively.

\subsection { Time dependent and regional values}In Fig. 1 (a-b) the number of
new firms appearing in each region is shown as  a function of (MC) time, (a) in
the case of a rather low $sel$, i.e. = 0.7 and in the case (b) of a rather high
$sel$ = 1.7. In Fig. 2 (a-b) the number of eliminated firms versus time, for the
above two values of $sel$ are shown. It is worth pointing here, the difference
with respect to the  case examined in ref. \cite{ACP}. In the latter paper the
long time stability (asymptotic value) was obtained through a sharp increase
followed by an exponential decay toward a $zero$ death (and also birth)
evolution. In the present case, the number of born or eliminated firms increases
smoothly and remains stable at a finite value. Recall that there are 8000 firms
(''possible parents'') at $t$=0. Observe the $x$-axis scale indicating the effect
of selection pressure on first time invasion of regions II and III.

A minimum in the birth and death number is always observed at short time (after
100 MCS) in region I. Notice from Fig. 1(b) and Fig. 2(b) that a strong $sel$
might quickly kill any evolution process : zero-birth and zero death. It is also
clear from comparing Fig. 1  and Fig. 2 that there is a rapid increase in
concentration in the newly opened regions as soon as the barrier at $L_{x1}$ is
removed. The death and birth processes are always quantitatively equivalent,
indicating that  a detailed balance equilibrium is intrinsically met in the
model.

The number of firms present in the three regions vs. time (in MCS) is shown for
several $sel$ values, from 0.4 till 1.7 in Fig.3 (a-d).  At very low $sel$ the
number of firms rather quickly reaches a rather high saturation value in each
region, in fact approximately equal to the initial value in region I.  At high
$sel$, Fig. 3 (c-d),  there is a marked drop in region I, while regions II and
III are invaded much later than in Fig.3 (a); the saturation level has  a lower
value than  the initial starting one in region I. The asymptotic number of firm
values is seen to decrease with $sel$. The strongly marked oscillations might be
considered as a signature of cycles \cite{Aoki2,Freeman,JMMALadek}. At high $sel$
values the effects are much enhanced. Also, regions II and III are invaded very
slowly and late in time. Yet the fitness optimal value in each region is easily
reached and remain stable in time (see Fig.4 below).

In Fig. 4(a-d) we show the average fitness $f$ in the three regions vs. MCS time
for several values of the selection pressure, ranging from 0.4 to 1.7.  Recall
that the values of the external field are respectively 0.3, 0.5 and 0.6, and the
wall at  $L_{x1}$ falls at $t$=100. For a low $sel$  (=0.4) small fluctuations
are seen around  the average (asymptotic) values which are close to 0.5 in all
regions. Since the selection pressure is low, the firms are not strongly forced
to being in perfect matching to the field value, whence large deviations from the
regional average fitness can be enforced.

For higher $sel$, data is quite scattered, and indicate a strongly hard learning
(adaptation) process, but the average fitness in region I is near 0.3, and that
in region III close to 0.6 as expected. For high $sel$ large fluctuations in the
average $f$ value are visible as in Fig. 4(b), at rather short time. At very
large $sel$ the beginning of the process is rather noisy, but the average fitness
stabilizes at a value expected from the imposed external field. We adopted scales
for the figures in order to show qualitatively the decrease in the variance of
the fitness distribution with time, for various selection pressures. The
adaptation is clearly faster in region I and II, and  the variance much reduced
as soon as the expected average values are  obtained.

Considering the  physical diffusion process, we recorded the maximum distance
reached by the right most firm as a function of time, and present it on a log-log
scale for different $sel$ values, from 0.3 till 1.8, in Fig. 5. Recall that due
to the lattice topology,  critical values are to be found at $x$= 50, 100 and
150. Studying Fig. 5, starting from high $sel$ values, it appears that a plateau
occurs (for $sel$ =1.8 and 1.5) near $L_{x2}$=100, also somewhat seen for $sel$ =
1.1.  The flow into region II and III seems more continuous at small $sel$
values. The small $tails$ sticking upwards (above $x$= 50) for $t> 1000$ witness
unsuccessful attempts to colonize regions II and III, suggesting that a
$sel$-dependent nucleation-growth process exists in the model. Moreover, for high
$sel$ values it appears that the region near the border at $L_{x1}$= 50  is
sometimes depleted., indicating a gain that high $sel$ values might destroy the
population at all. From Fig. 5 (and others not shown here) we can calculate the
time necessary for the first firm to reach the region right boundary, i.e.
respectively $L_{x2}$=100  and $L_{x3}$ = 150, as a function of the selection
pressure (Fig. 6). The curves look like high order power laws (with exponents
equal ca. 4 and 5 respectively), but more complicated functions might be tried.

\subsection { Asymptotic values}

It is of interest to observe beyond the transient regimes the equilibrium values.
In Fig.7 (a)  the normalized asymptotic values of the born firms in the three
regions are shown as a function of $sel$. The ''normalization'' is made with
respect to the total asymptotic number of firms in the region shown in Fig. 7(b).
''Asymptotic'' meaning here that we have averaged the last 500 entries for  10
000 MCS runs. The same type of data is obtained $quantitatively$ for the number
of disappearing firms (not shown). A maximum occurs near $sel$ = 0.8 with
different analytic behaviors on both sides of this value. The increase goes like
a stretched exponential as for diffusion-controlled growing entities\cite{Avrami}
at low $sel$ ($<0.8$), i.e. $ sel^{0.45} exp(-0.5/\sqrt(sel))$, but the decay at
high $sel$ has been found to be  markedly a power law $sel^{-0.40}$, like for
critical systems \cite{StanleyRMP}. Recall that each region may contain at most
10 050 entities, a number to be compared to the number of asymptotically existing
firms, as displayed on Fig. 7 (b). For completeness, the asymptotic fitness in
the three regions is shown vs. $sel$ in Fig. 8. This indicates that only for
large $sel$ ($>0.8$)  has the asymptotic average fitness almost reached the
expected one from the field conditions.

\subsection { Regional  behaviors}

We show the short time dependence  ($t$ $\leq $800 MCS) of  the (average over a
''vertical column'') firm concentration vs.  the column number ''$x$'' for rather
strong (1.3) and very strong (1.7) $sel$ values in Fig. 9. A drop in the overall
concentration is well seen in region I at short time, but the concentration
recovers later on, as seen from previous paragraphs and figures. For $sel$ =1.8
the selection pressure is too high to maintain any entity in region I after a few
MCS.

The dependence of the firm (vertical) concentrations on position ''$x$'' at
different (longer) times is shown in Fig. 10 (a-c) for $sel$ values just above
the $critical$ one, i.e. $\simeq$ 0.8.

Interesting  and unexpected  features occur. At very low $sel$ (not displayed) a
saturation level is reached right from the beginning of the simulation,
indicating ''easy adaptation'' everywhere. At high $sel$ the third region is
invaded $ca.$ after 5000 MCS,  while  region II is invaded after 800 MCS. We
emphasize the occurrence of large well marked ''oscillations'' in the local
concentrations. Dips do occur, see Fig. 10 near the region borders, - likely due
to the difficulty for firms to reach the most appropriate fitness together with
their neighbors, due to the interregion field gradient.

Finally  the  (vertically averaged) fitness $vs.$ position along the $x$-axis
taken at three time steps  and for two $sel$ values on both sides of the critical
one are shown in Fig. 11. Notice the different time(s) chosen for such snapshots.
For medium $sel$ (=0.7), the scatter around the expected theoretical fitness
imposed by the external field is not large but the average values nevertheless do
not closely correspond to the expected ones from the imposed field, in all three
regions, - the more so on the average , and asymptotically as seen above. At high
$sel$ (=1.7), the data is more scattered but the expected values are better
recognized as fulfilling the external field condition.

\section{CONCLUSION}

The dynamics of an economic firm population has been considered through a model
considering entities characterized by a scalar number, diffusing on a regular
lattice, merging, collapsing or creating spin-offs. The economic environment is
described by a static regionally dependent field. The adaptation and evolution of
the firms to the field condition have been mimicked by a Darwinian-like selection
rule but with a fitness evolution equation for the newly appearing firms taking
into account that of the parents. The firm fitness distribution is shown to have
some best adaptation difficulty, as discussed by Yamasaki et al. \cite{Kazuko}.
We have found an unexpected and interesting feature: the local changes of the
environment is leading to sharp variations, almost discontinuous ones, in the
fitness and concentrations, in particular when the field gradient is strong.  A
lack of self-organization is thus seen at region borders. This is due to the fact
that for an entity attempting an adaptation, the learning process is quite hill
climbing as in NK models\cite{Kauffman1998}. Indeed the evolution of an economy,
in which the functioning of companies is interdependent  and depends on external
conditions, through natural selection and somewhat random mutation is similar to
bioevolution on NK fitness landscapes as described by
Kaufmann\cite{Kauffman1998}. In these evolutionary economics
models\cite{Frenken}, economic agents randomly search for new technological
design by trial-and-error and run the risk of ending up in sub-optimal solutions
due to interdependencies  between the elements in a complex system. As argued by
Frenken\cite{Frenken} these models of random search are legitimate for reasons of
modelling simplicity, but remain limited as these models ignore the fact that
agents can apply heuristics.'' We totally agree. Indeed there is no mimicking nor
endogeneous search for optimisation as in
\cite{Kauffman2000,Auerswald2000,Frenken} in the present model,  the algorithm
being only geared toward  reducing the distance between the company fitness and
the external field value.

One might take into account in a more proper way the number of business contacts,
the evolving firm neighbors : the  development of a market or a business plan, or
a macroeconomy depends on the available number of contacts (bonds between nodes
on a network)  whence to the creation and destruction of contacts. A large stock
of relational capital (business contacts) usually increases the sold output of a
representative firm. Other lattices or network structures should be usefully
studied.

It would be of interest to consider specific (historical) cases, and connect the
MC simulations time scales to real cases. The time scale of changing environments
is an available  parameter which could indicate how robust the  model can be with
respect to realistic  adaptation and evolution. Some flexibility for defining
time scales exist in the (i) $b$ and (ii) $sel$ values: (i) The value of the
spin-off creation probability might be too large for previous economies. It has
been really shown in the line of the ACP model that for the best fitted company a
small $b$ value leads to chaotic behavior while a large $b$ leads to a monopoly
like situation\cite{JMMALadek}. (ii) A $sel$ (space and time dependent) value
could be found by examining macroeconomy data. A sort of $critical$ selection
pressure exists separating different birth-death regimes. This might be used as a
time and field scale definition when counting bankruptcies in countries,
including per industry type.

Contrary to the Bak-Sneppen model \cite{BakSneppen}, the average fitness(es) or
concentration(s) do not seem to appear as power laws (like those characterizing
avalanches), except in the high selection pressure regime and for long
(asymptotic) times. The power law dependence, which is a usual signature of
self-critical organization is only seen for a high selection pressure in the
asymptotic behavior of the birth-death number. Nevertheless from the set of
results, as presented here above, we observe that  in such a Darwinian evolving
economic world: (i) there are relatively well marked effects due to the
"selection pressure"; (ii) temporal field changes can imply a stable density
distribution, as in Ausloos {\it et al.}\cite{ACP}; (iii) a diffusion process
together with a business plan, and the selection pressure smoothly leads to
asymptotic equilibrium states with respect to birth and death processes; (iv) the
fitness does not always reach the {\it a priori externally} imposed field value;
(v) the role of the gradient (economic) field at borders might indicate either
complex oscillations or chaotic behaviors in such regions.

We are aware that further improvements  are needed. We are drastically
caricaturing macro and micro economy field conditions, as well as the description
of the "internal" interactions sequence(s).   Surely a company, or a set of
industries, should not be described by one scalar number $f_i$, but rather a
vector (or matrix) model coupled to a (so called external) vector (or matrix)
field should be examined. Moreover the birth and death process description
through merging and spin off's could also be stochastic or evolutive. The spatial
distribution in the distinct regions might be also on interest\cite{Aoki3} as in
biology\cite{AMPV}.

\vskip 0.5cm {\bf Acknowledgements}

\vskip 0.3cm MA  thanks Masano Aoki, Mich\`ele Sanglier, Mieko Tanaka-Yamawaki 
and Kazuko Yamazaki for stimulating discussions, references and comments.

\vskip 0.3cm

\section*{Figure captions} \vspace*{0.2cm} Fig. 1 --   Number of new firms born in  region I, II, III ( +, x, dotted square,
respectively) as a function of time, (a)  $sel$  = 0.7 ; (b)  $sel$ = 1.7

\vskip 0.6cm

Fig. 2 -- Number of eliminated firms  in the three regions versus time, for (a)
$sel$  = 0.7 ; (b) $sel$ = 1.7

\vskip 0.6cm

Fig. 3 --  Number of firms in the three regions $vs.$ time (in MCS)  for (a)
$sel$ = 0.4; (b) $sel$ = 0.9; (c) $sel$ = 1.3; (d) $sel$ = 1.7

\vskip 0.6cm

Fig. 4 -- Average fitness in the three regions I, II, III ( +, x, dotted square,
respectively)  vs. MCS time  for several values of the selection pressure (a) =
0.4; (b) = 0.9;(c) = 1.3; (d) = 1.7

\vskip 0.6cm

Fig. 5 -- Maximum distance reached by the right most firm as a function of time,
on a log-log scale for different $sel$ values

\vskip 0.6cm

Fig. 6 -- Shortest time necessary for a firm to reach the region II (100) or III
(150) right boundary, i.e. $L_{x1}$=100, $L_{x2}$=150, as a function of the
selection pressure ; curves indicate the best power law fit, i.e. $ca.$ 4 (top)
and 5 (bottom) respectively

\vskip 0.6cm

Fig. 7  -- (a) Normalized asymptotic values values of the born firms in the three
regions  as a function of various $sel$;  ''Normalization'' is with respect to
(b) total asymptotic number of firms in the respective region

\vskip 0.6cm

Fig. 8. -- Asymptotic fitness in  regions I, II and III   $vs.$ $sel$  ; recall
that the field values are imposed to be 0.3, 05. and 0.6 respectively

\vskip 0.6cm

Fig. 9 -- Short time dependence of the firm concentration in regions I, II, III
vs. spatial coordinate ''$x$'' for   $sel$ values (a) = 1.3, (b) = 1.7 and at
different time moments - 10,20,50 and 800 MCS.

\vskip 0.6cm

Fig. 10 -- Dependence of the firm concentrations on position ''$x$'' at different
times  for $sel$ =  (d)   1.1;  (e) 1.5; (f) 1.7

\vskip 0.6cm

Fig. 11 -- Vertically averaged fitness $vs.$ position along the $x$-axis at three
time steps    and for two $sel$ values, i.e. (a) 0.7 and (b) 1.7

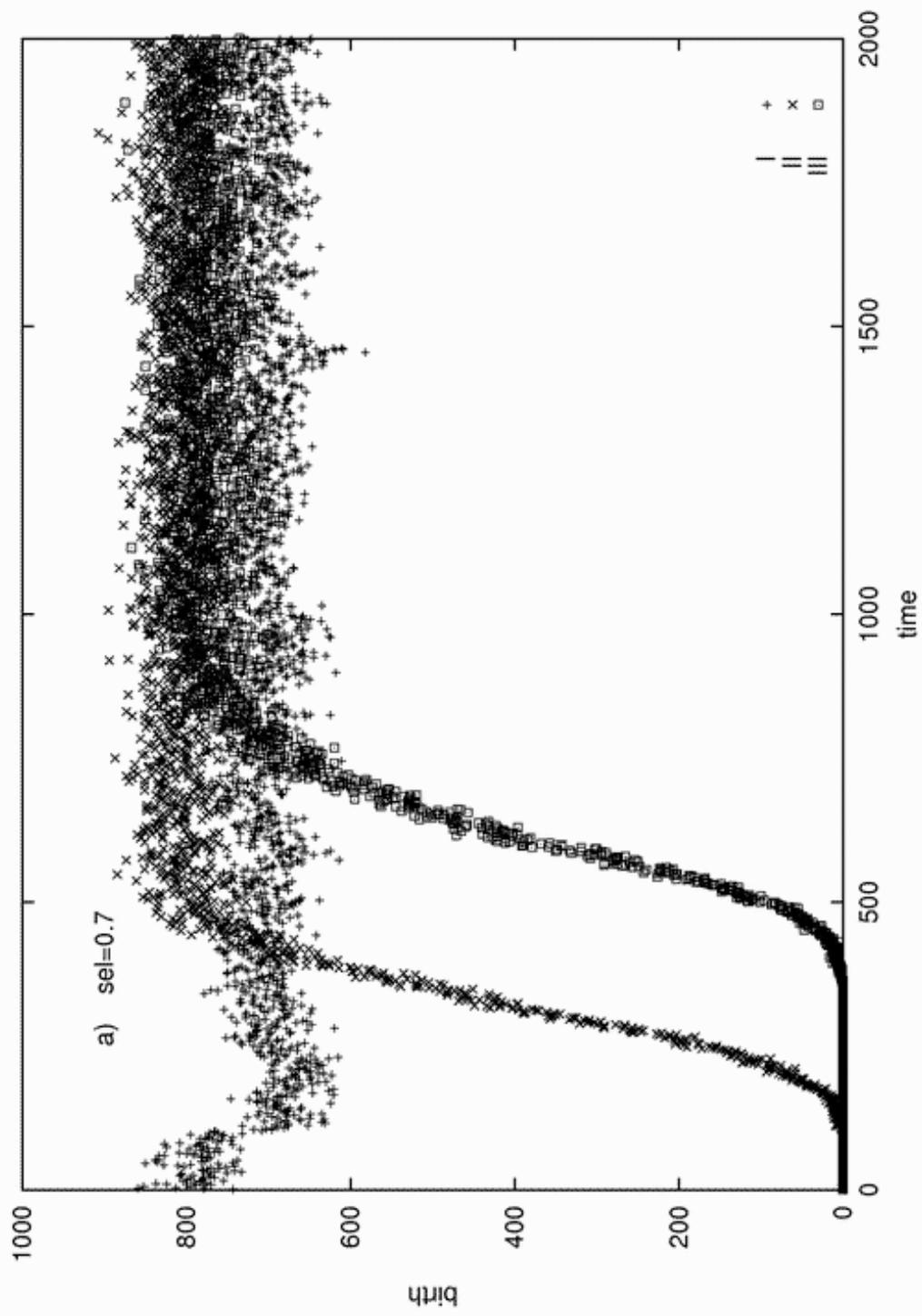
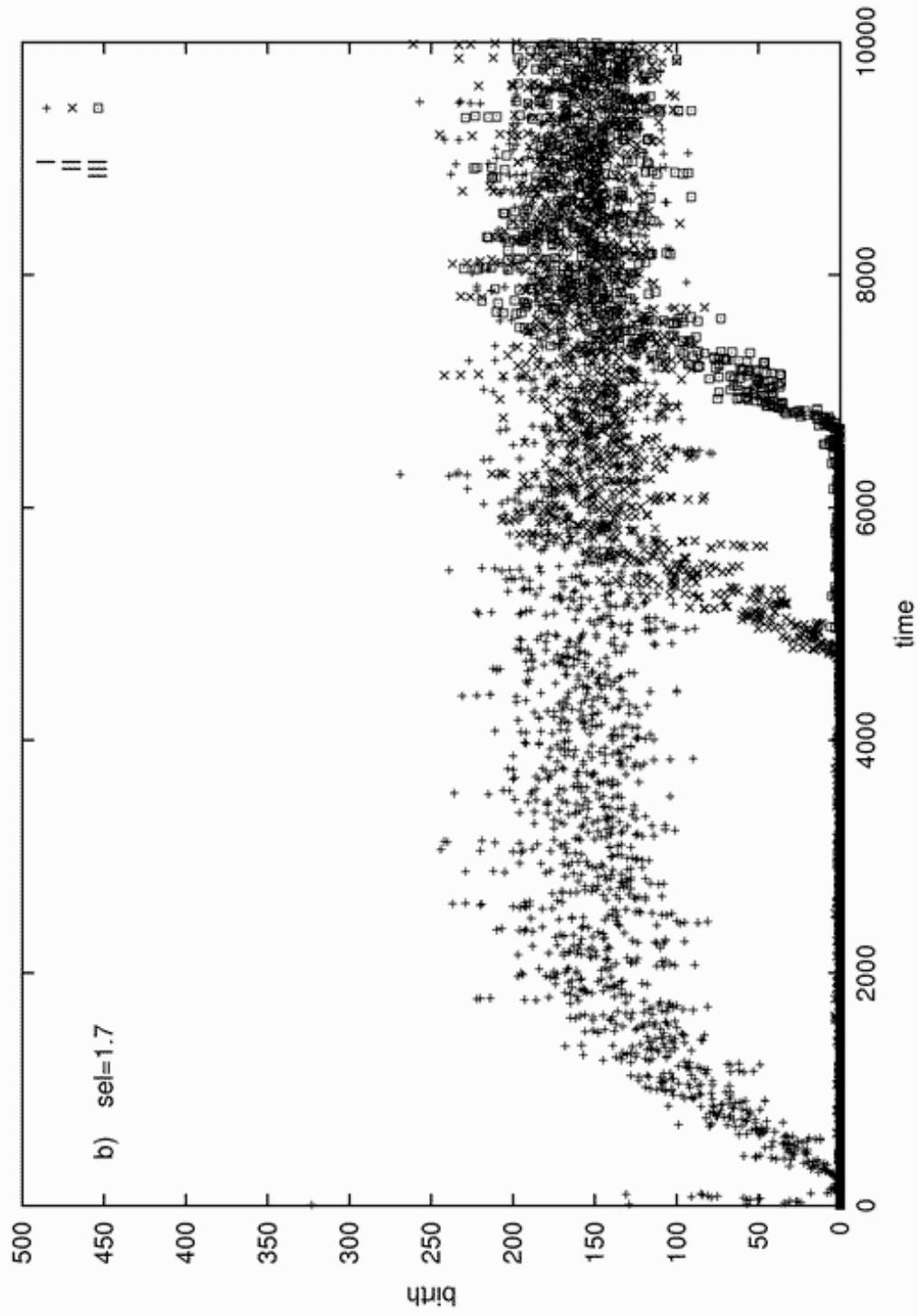
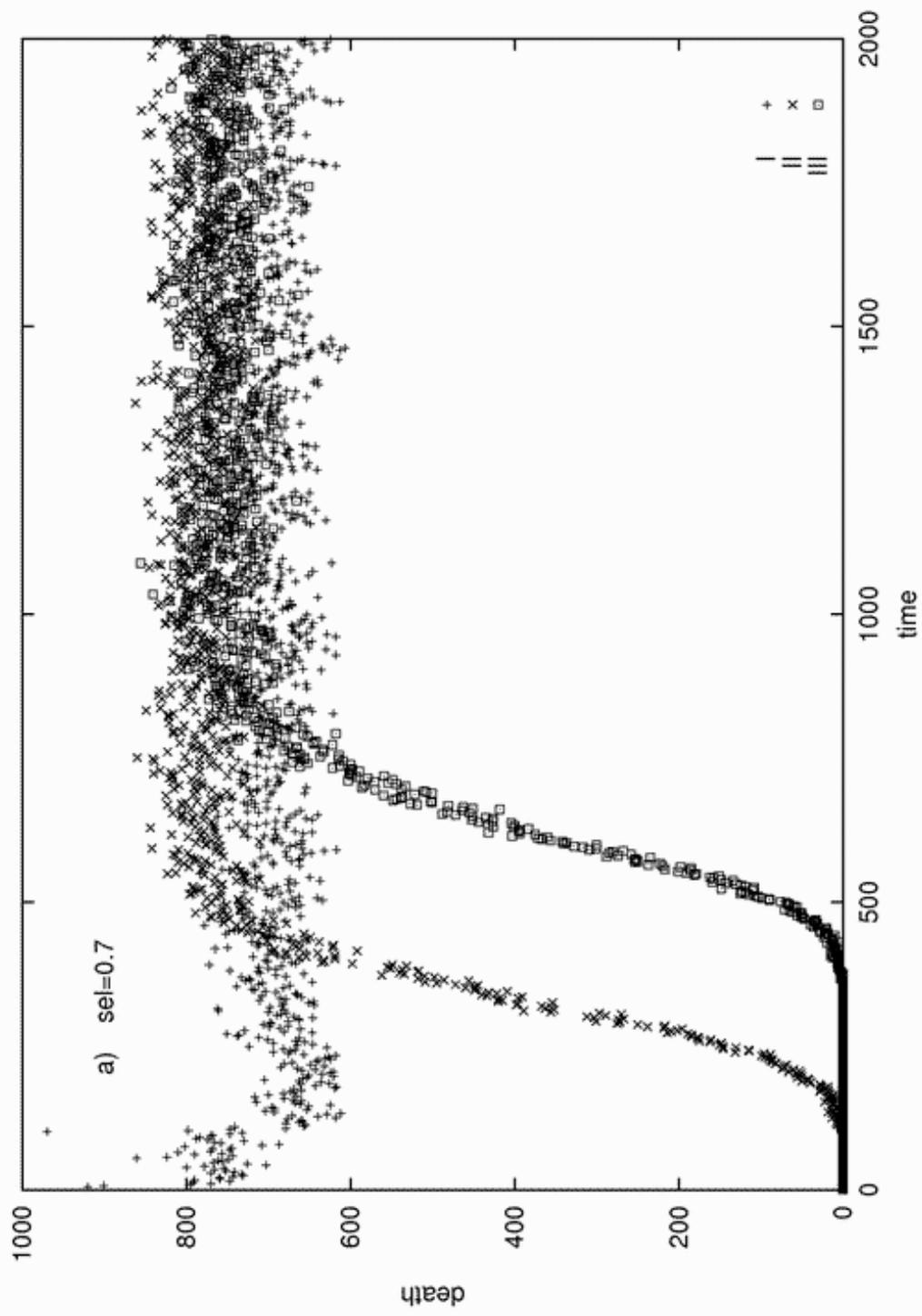
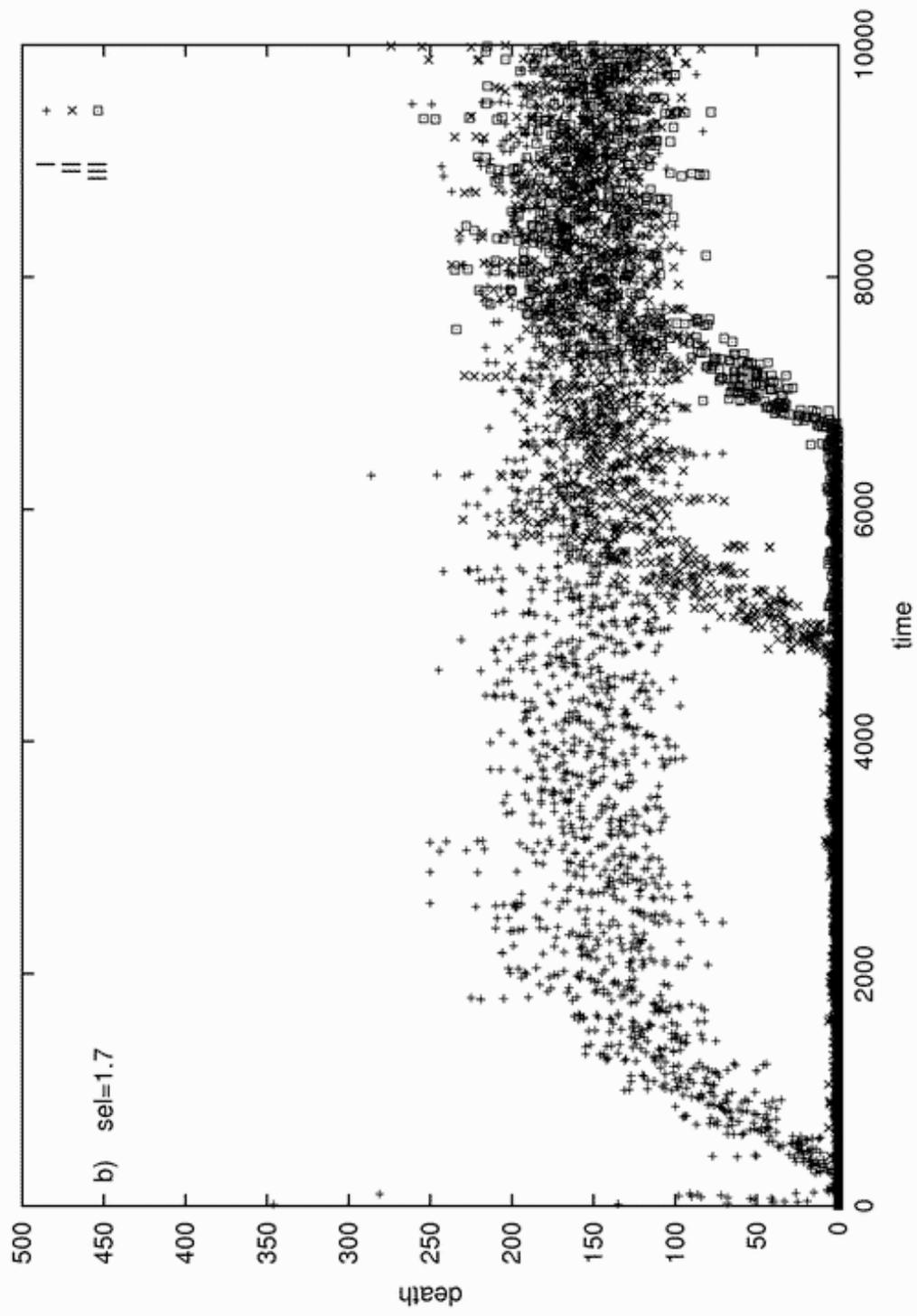
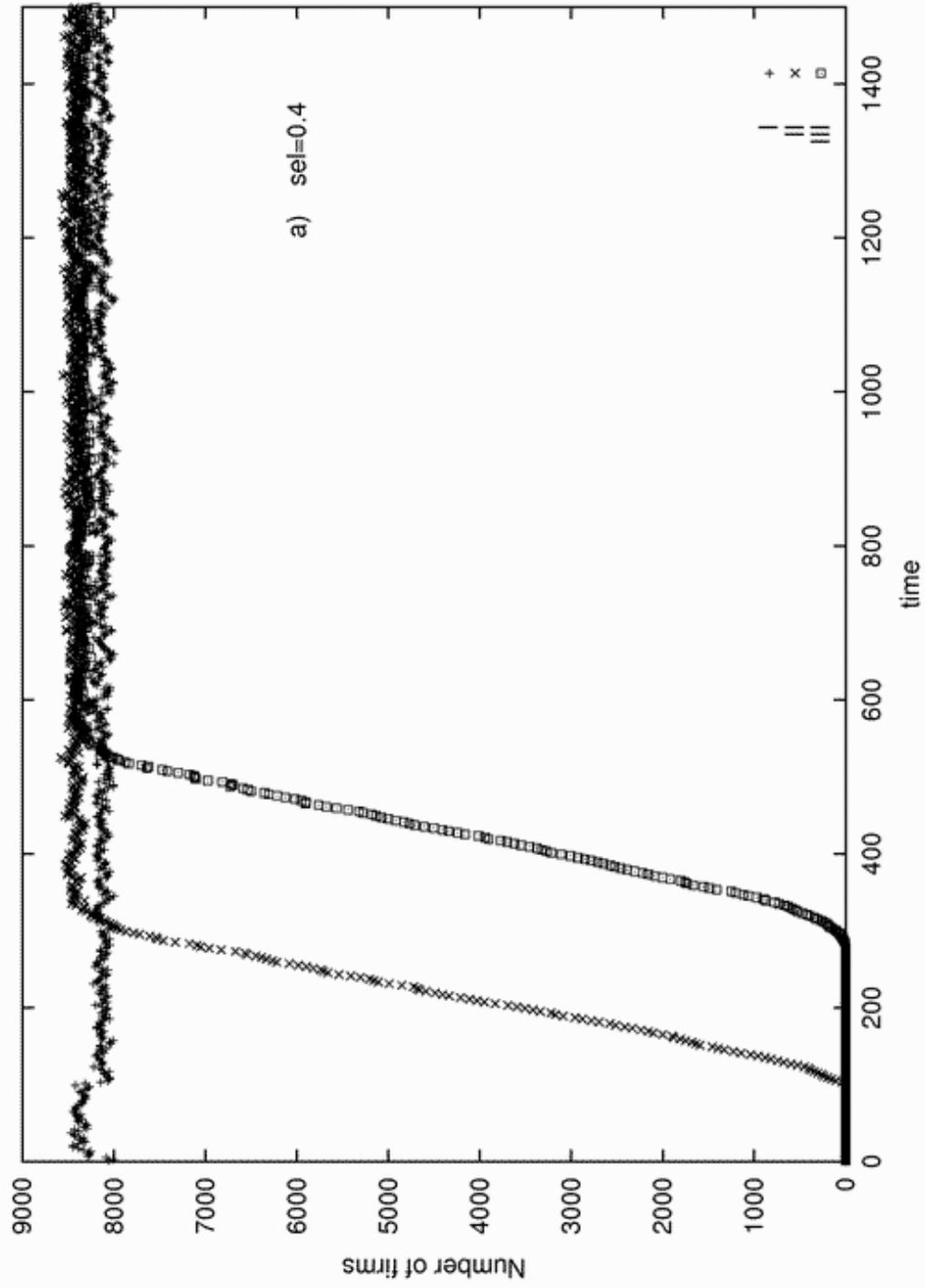
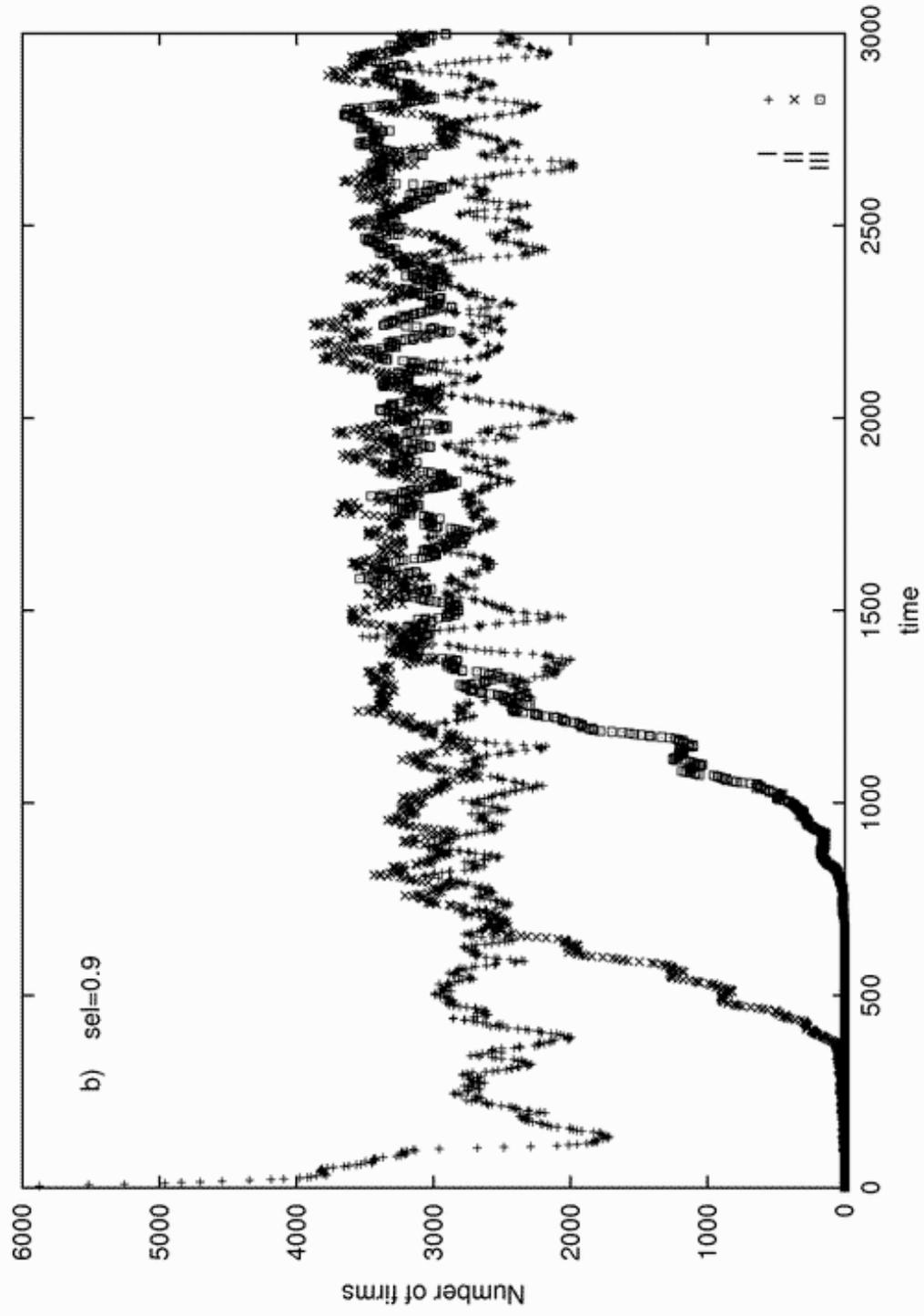
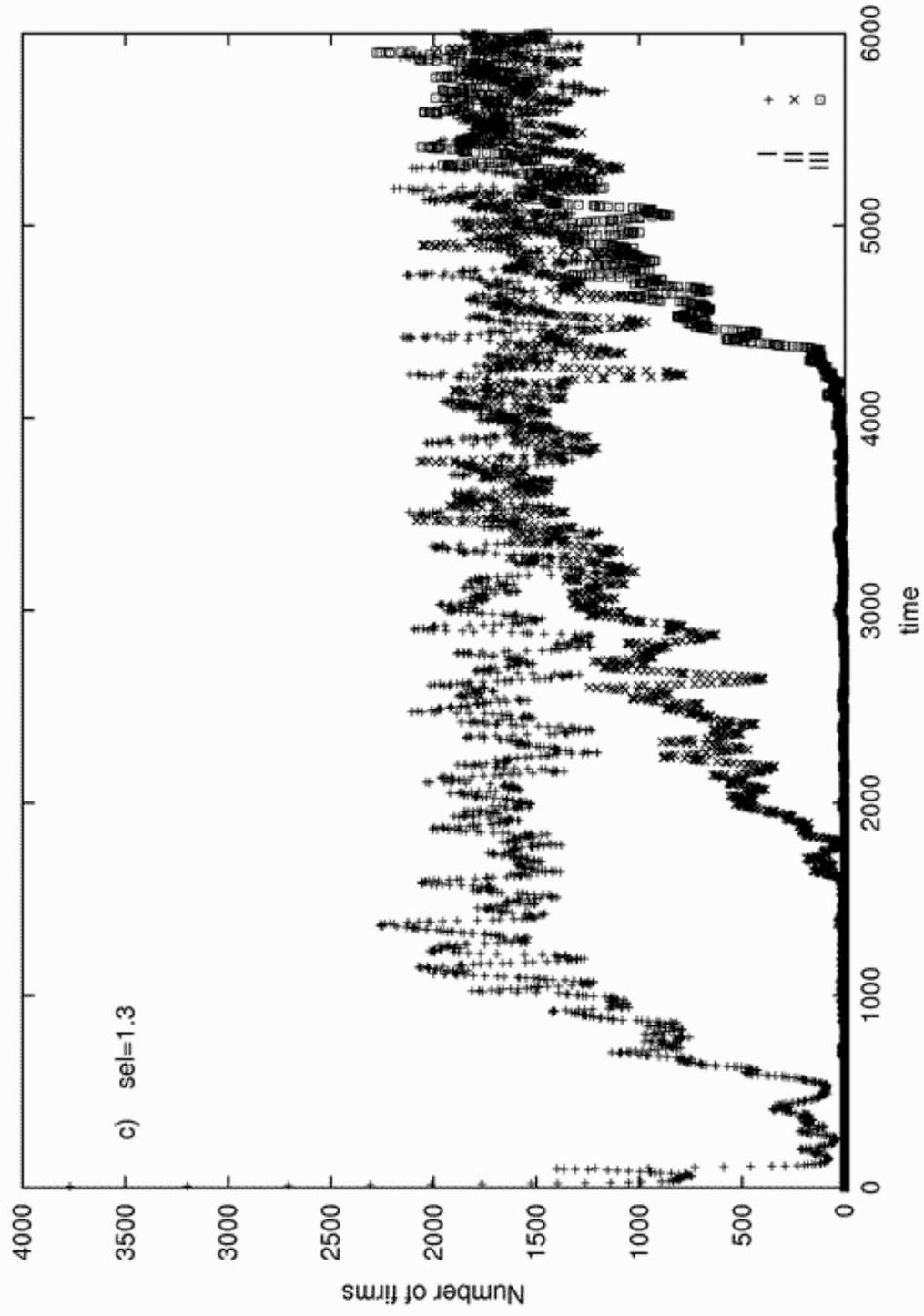
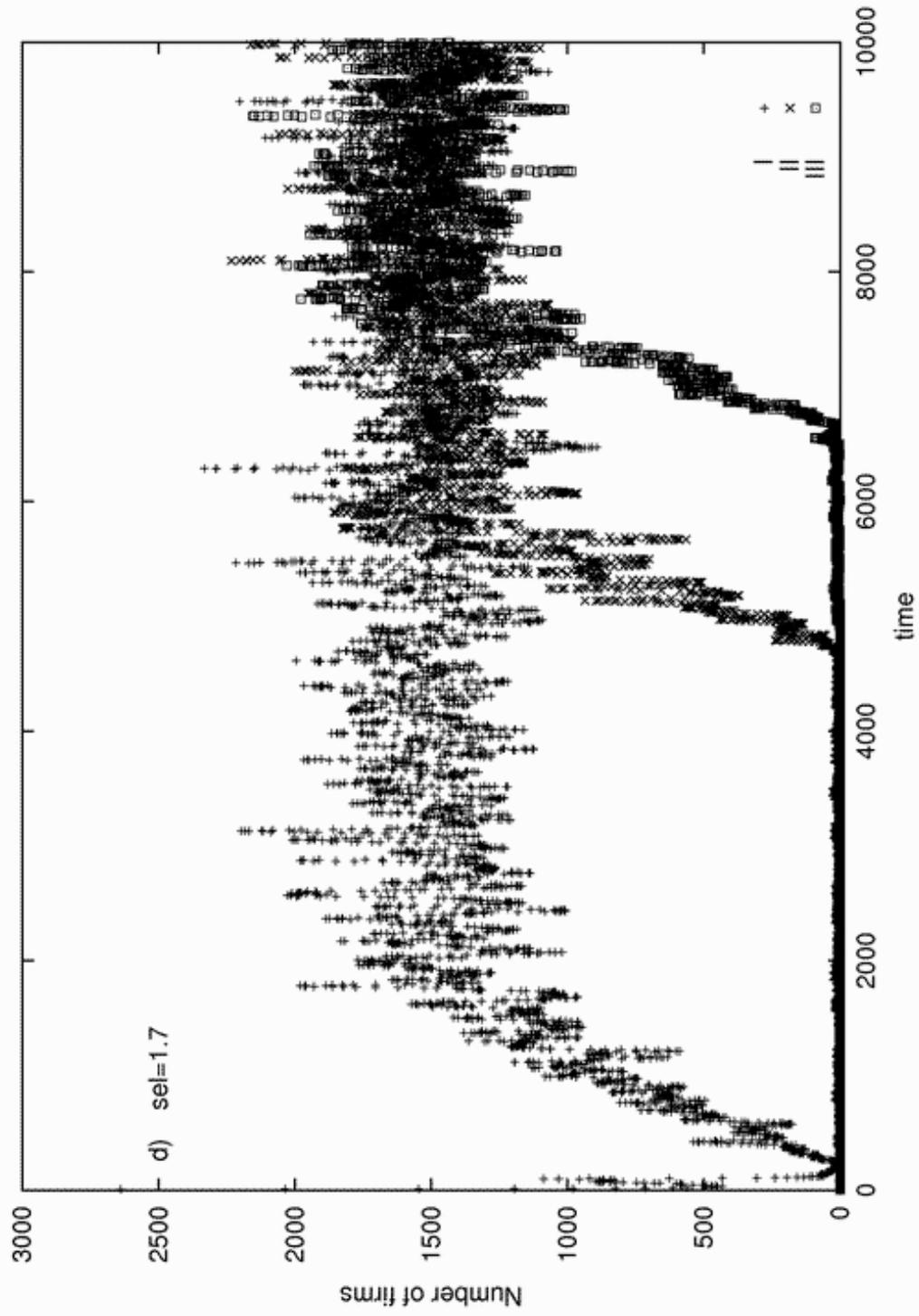
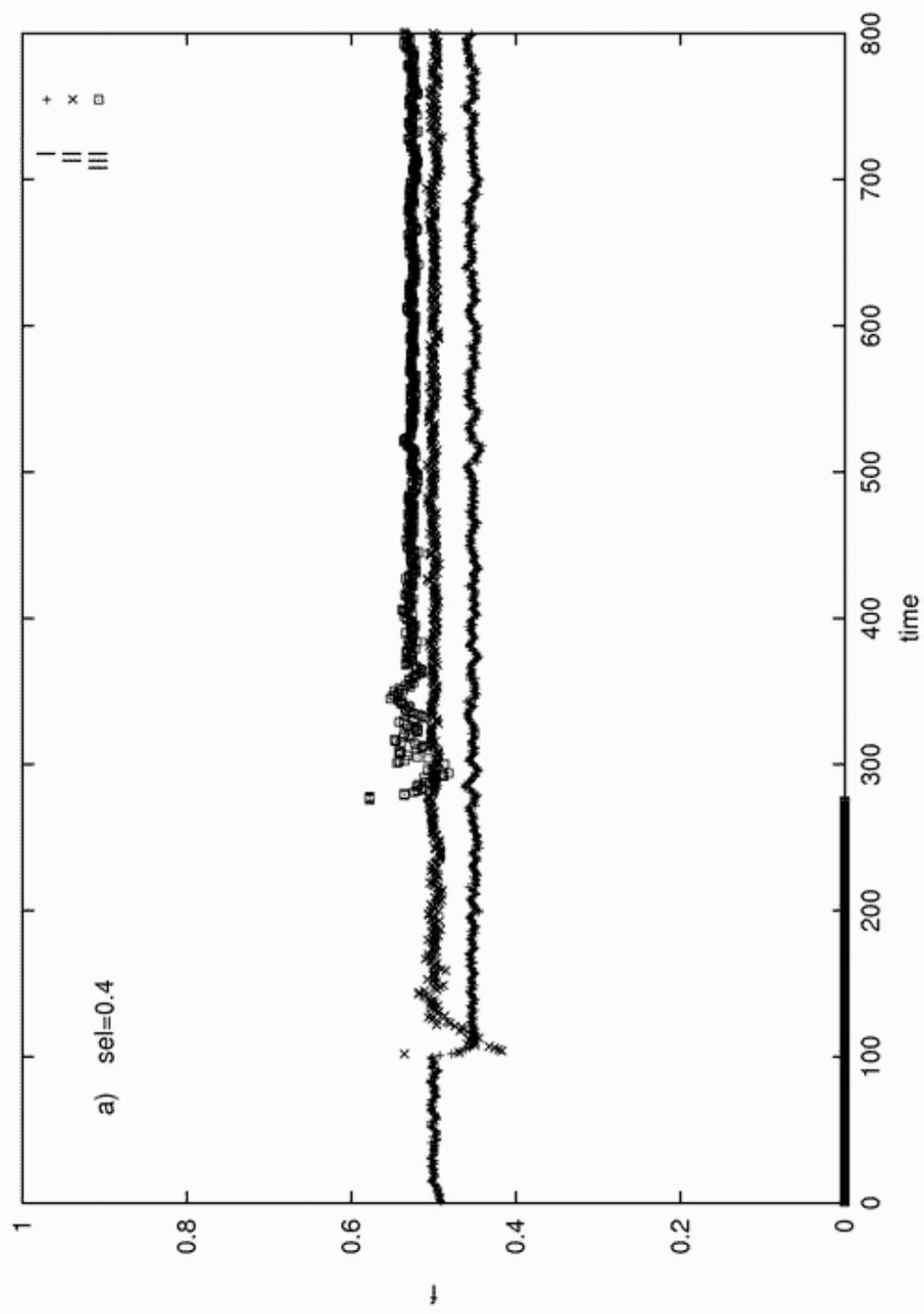
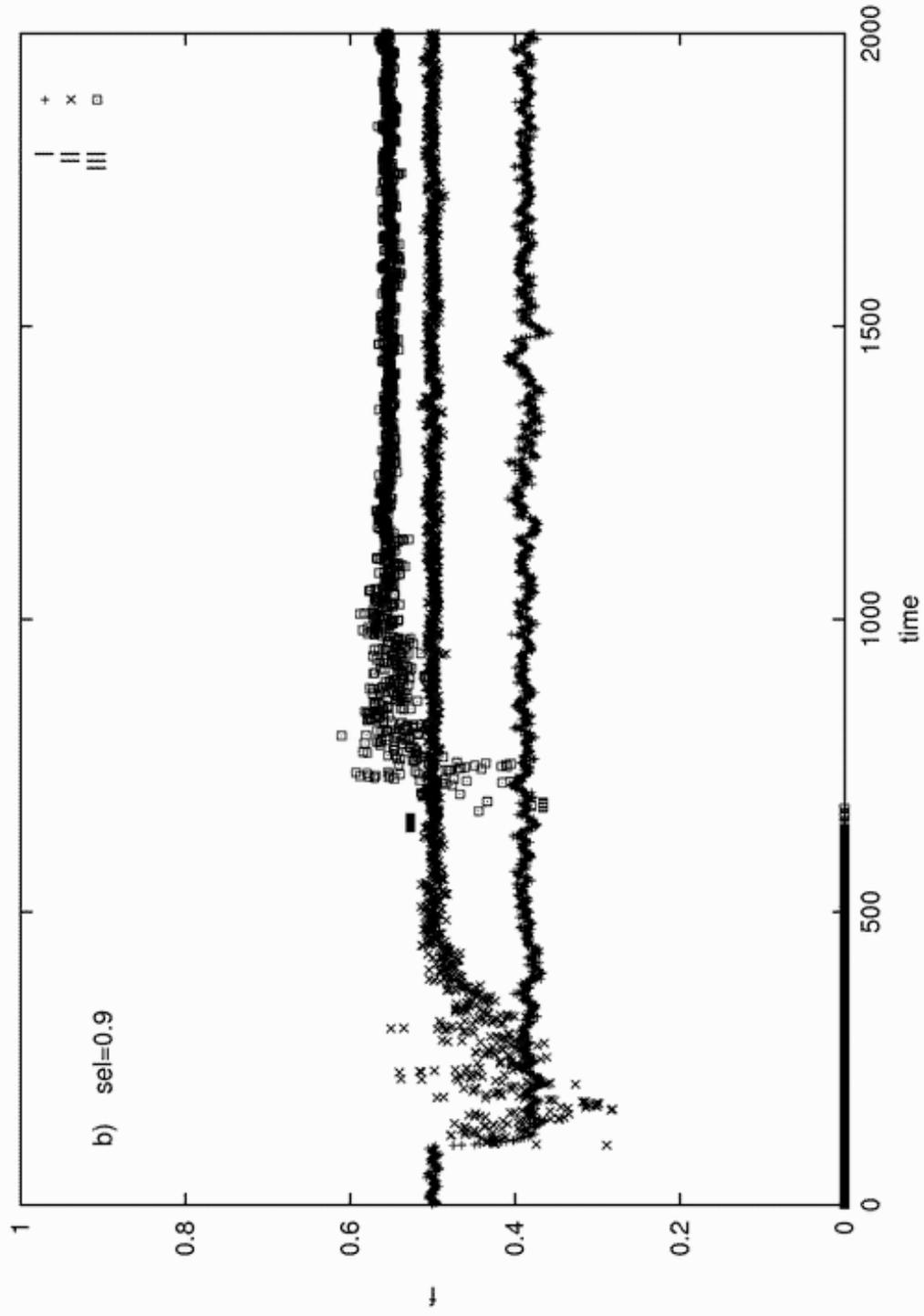
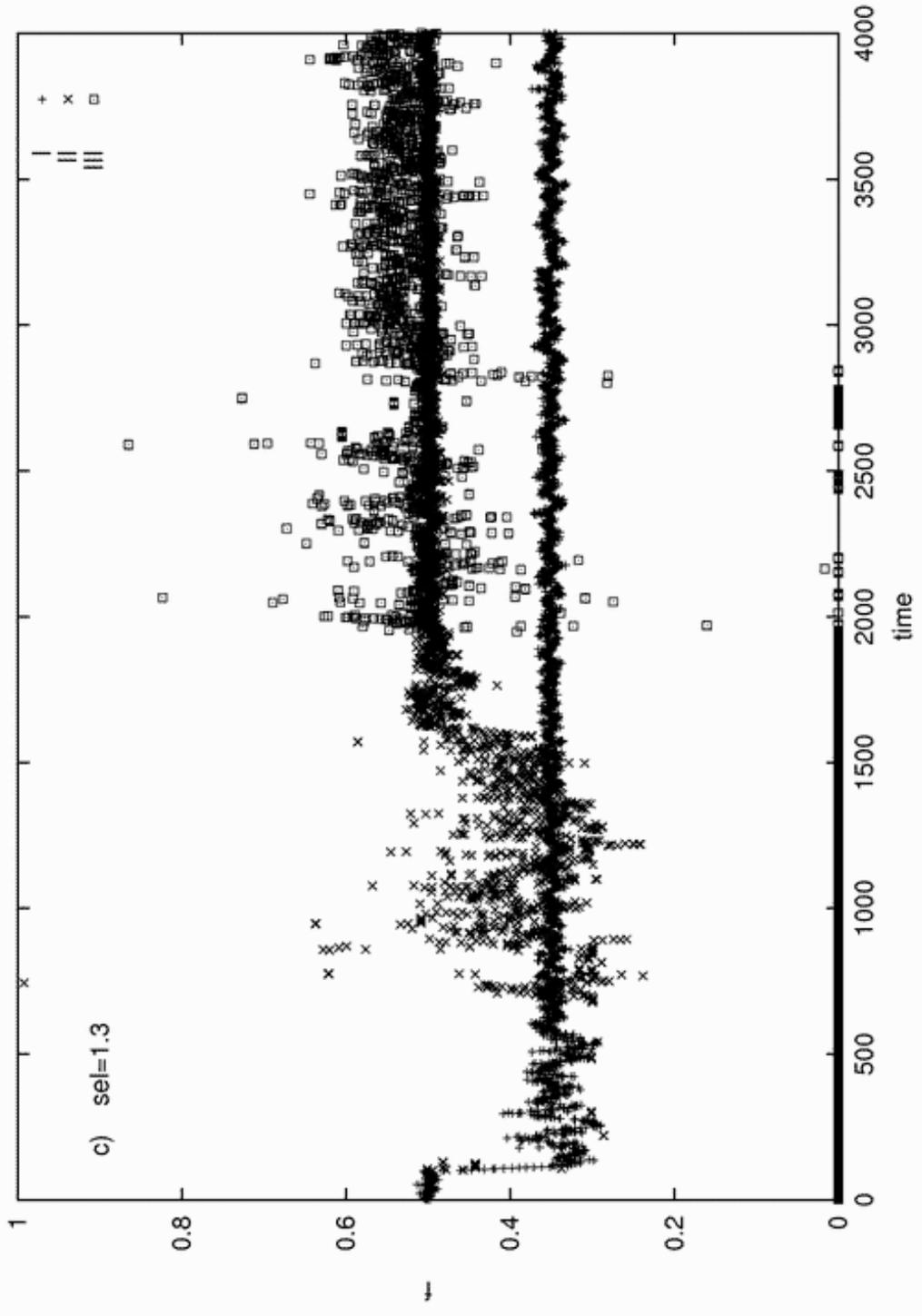
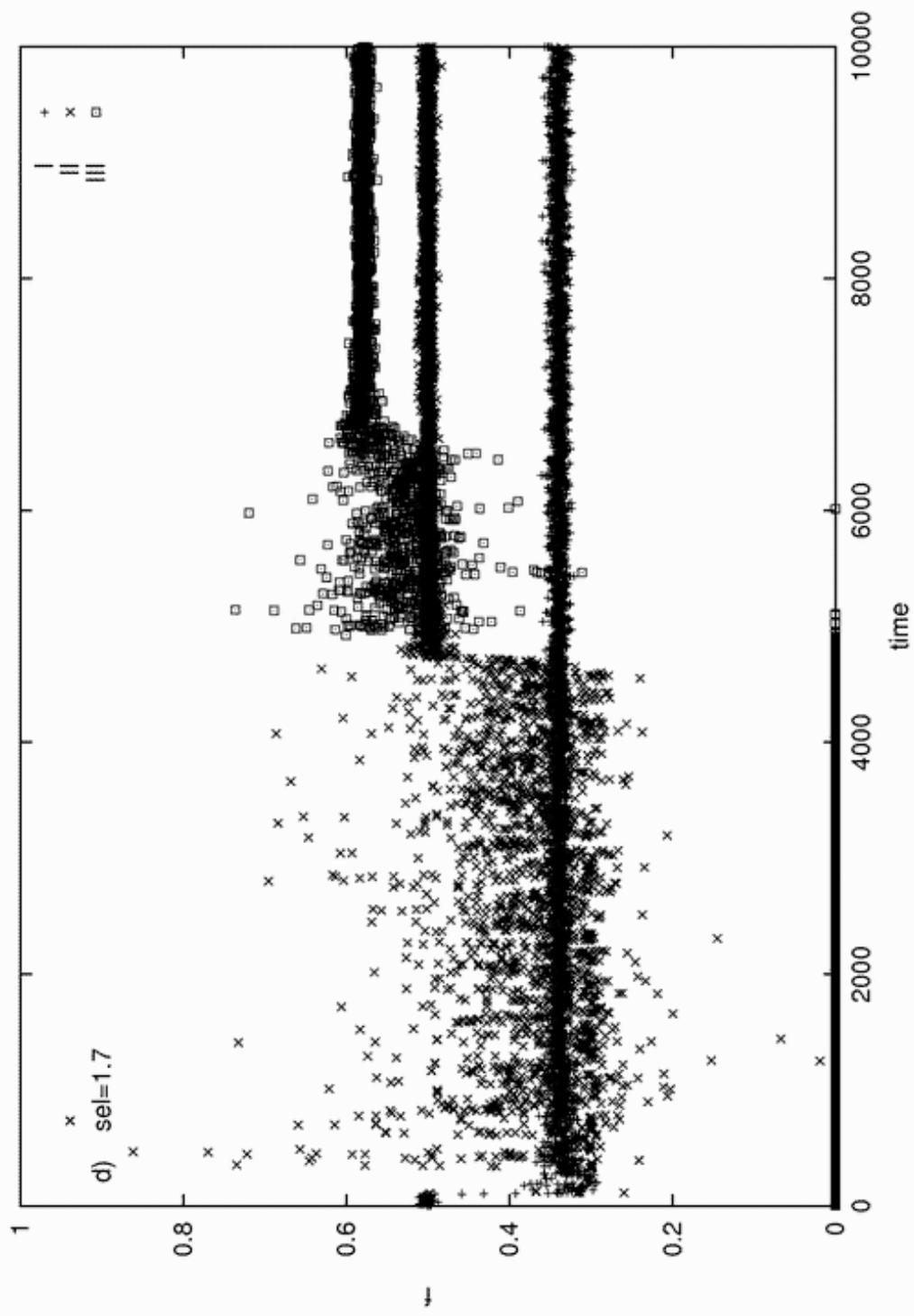
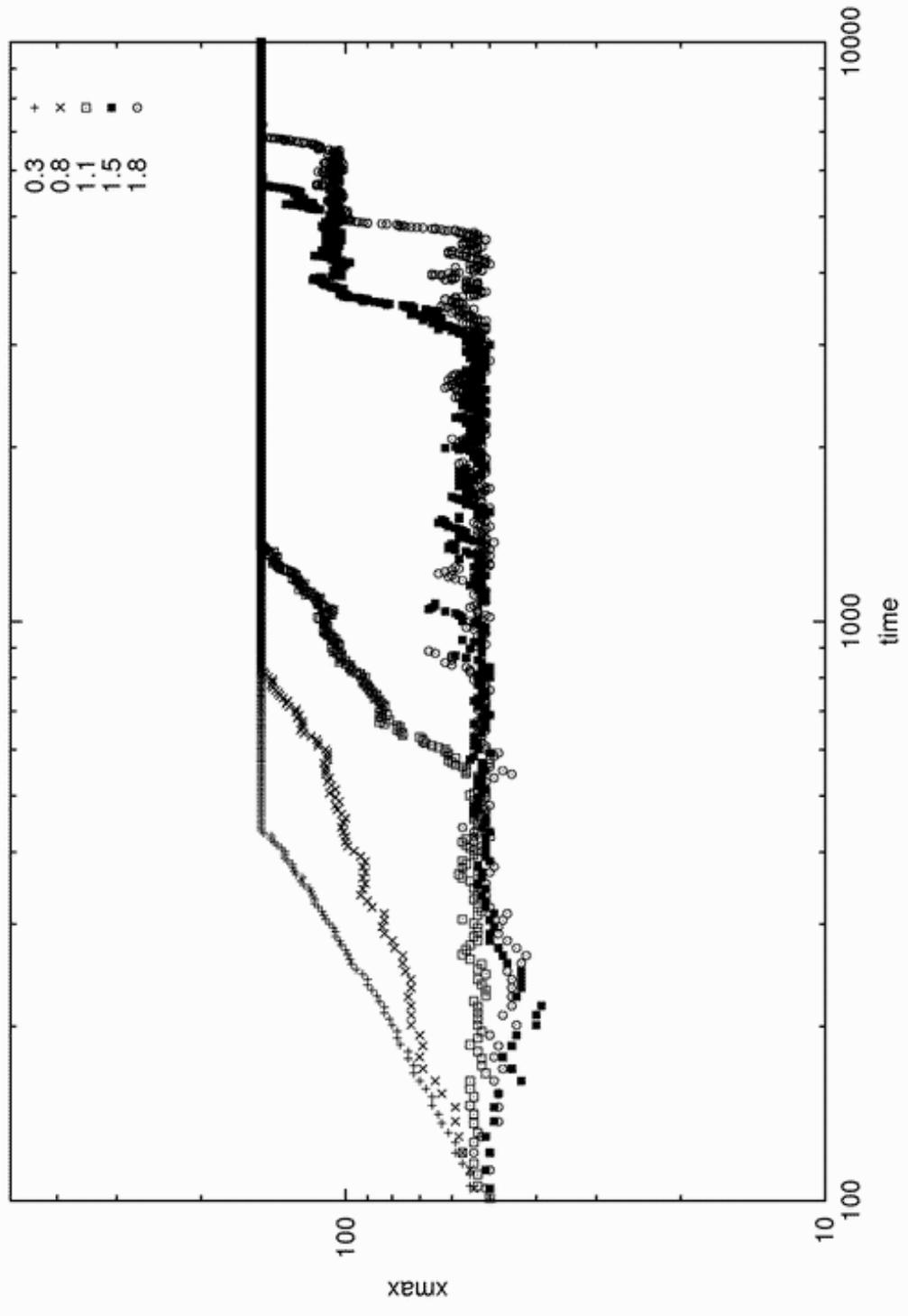
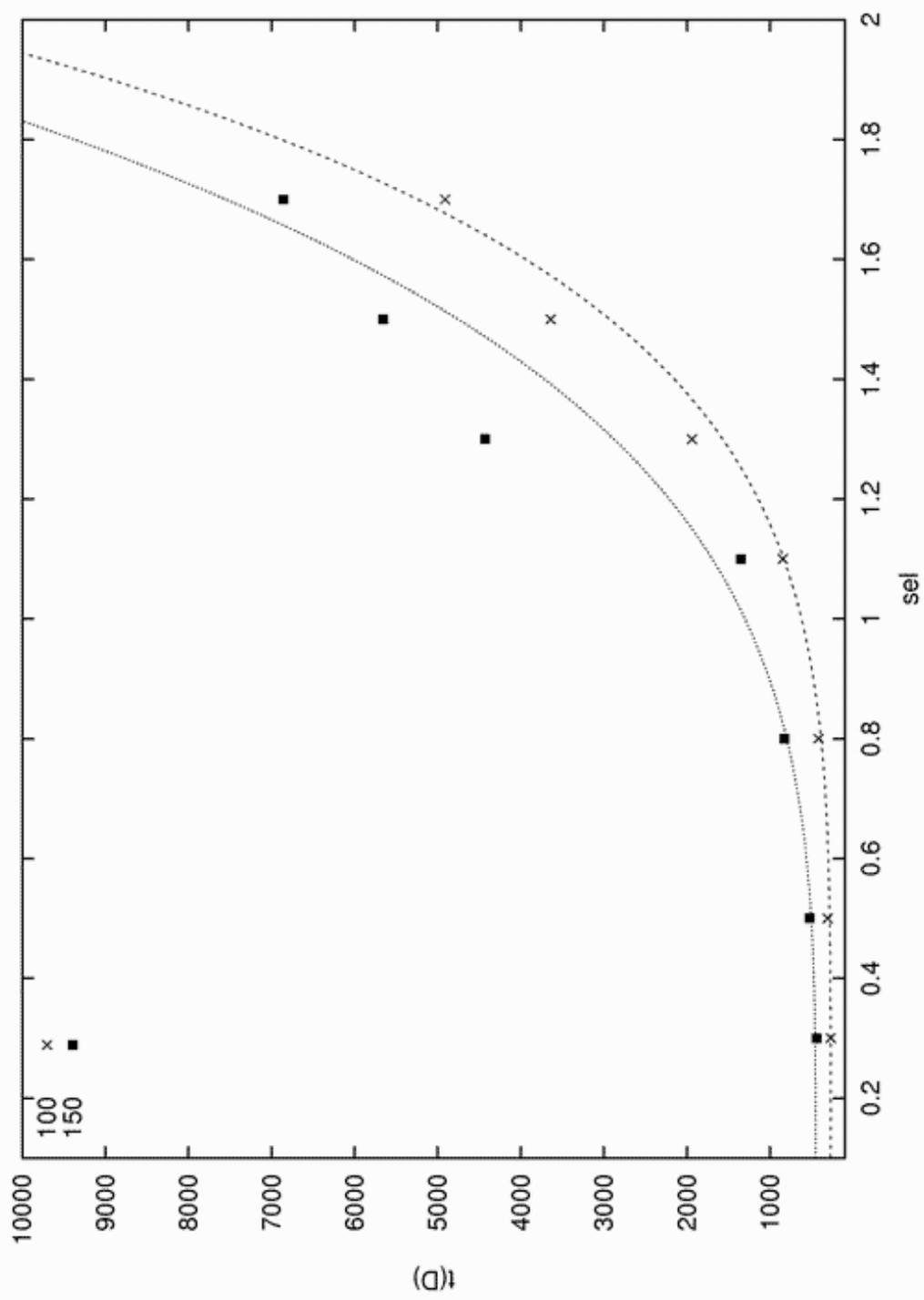
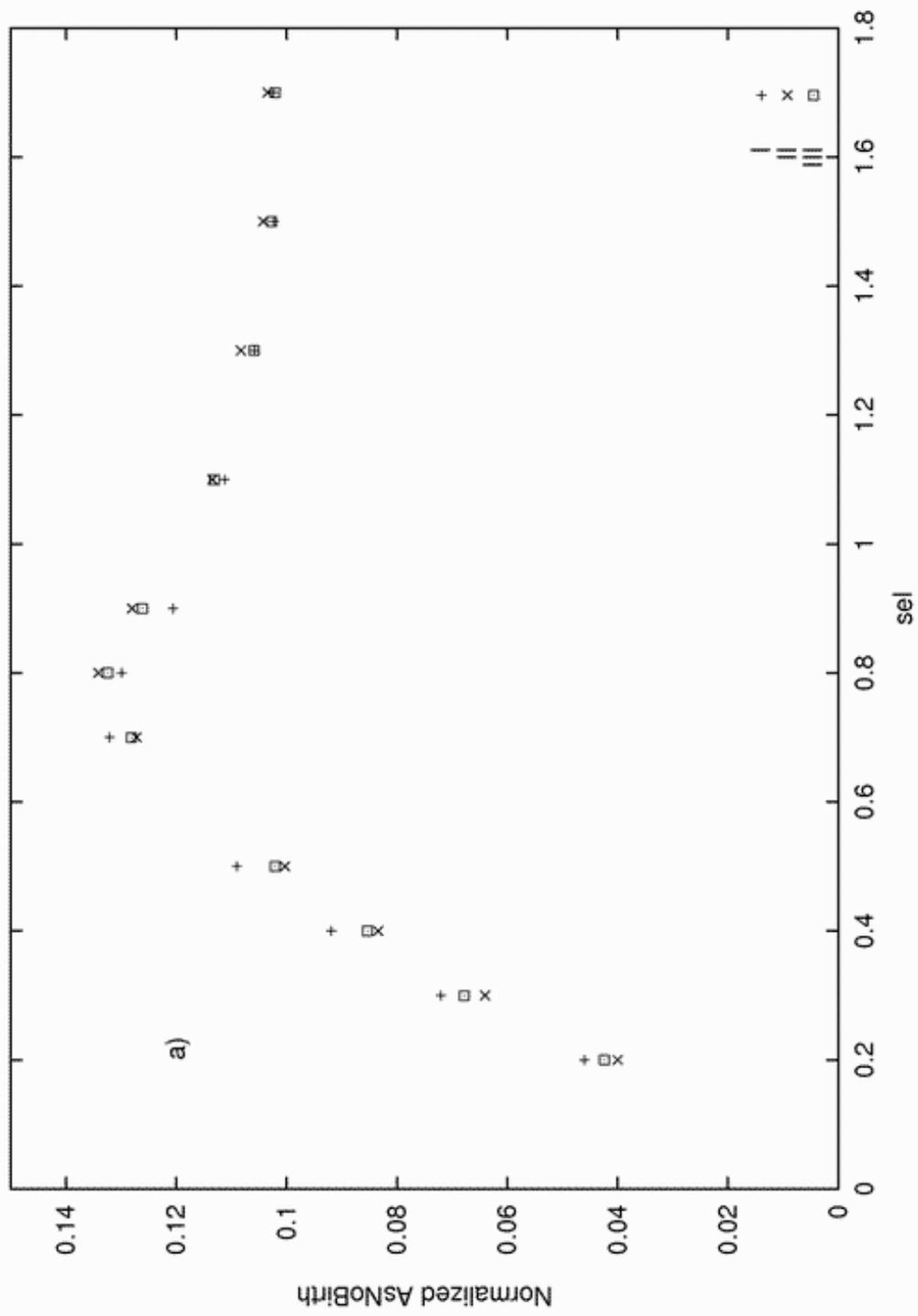
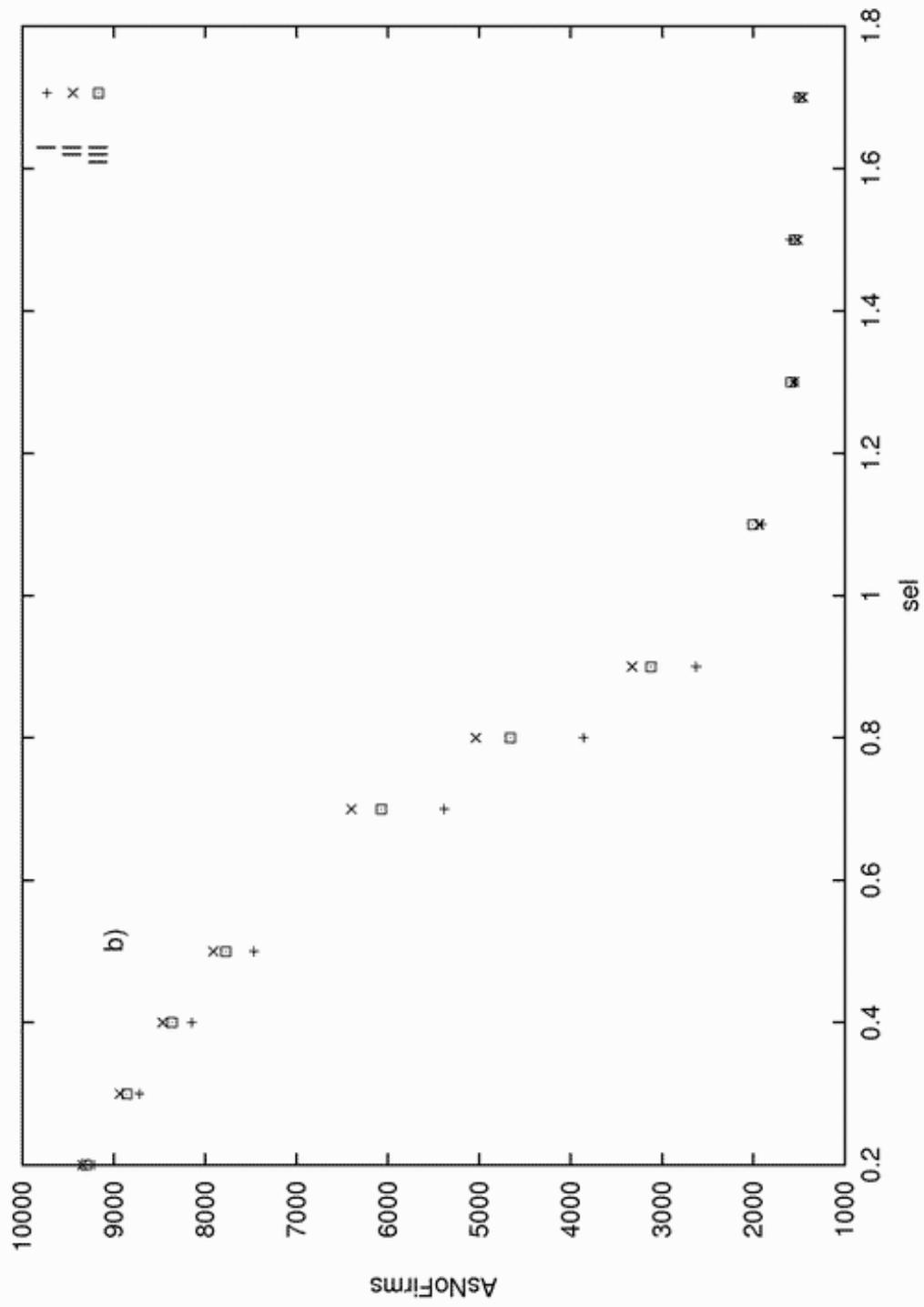
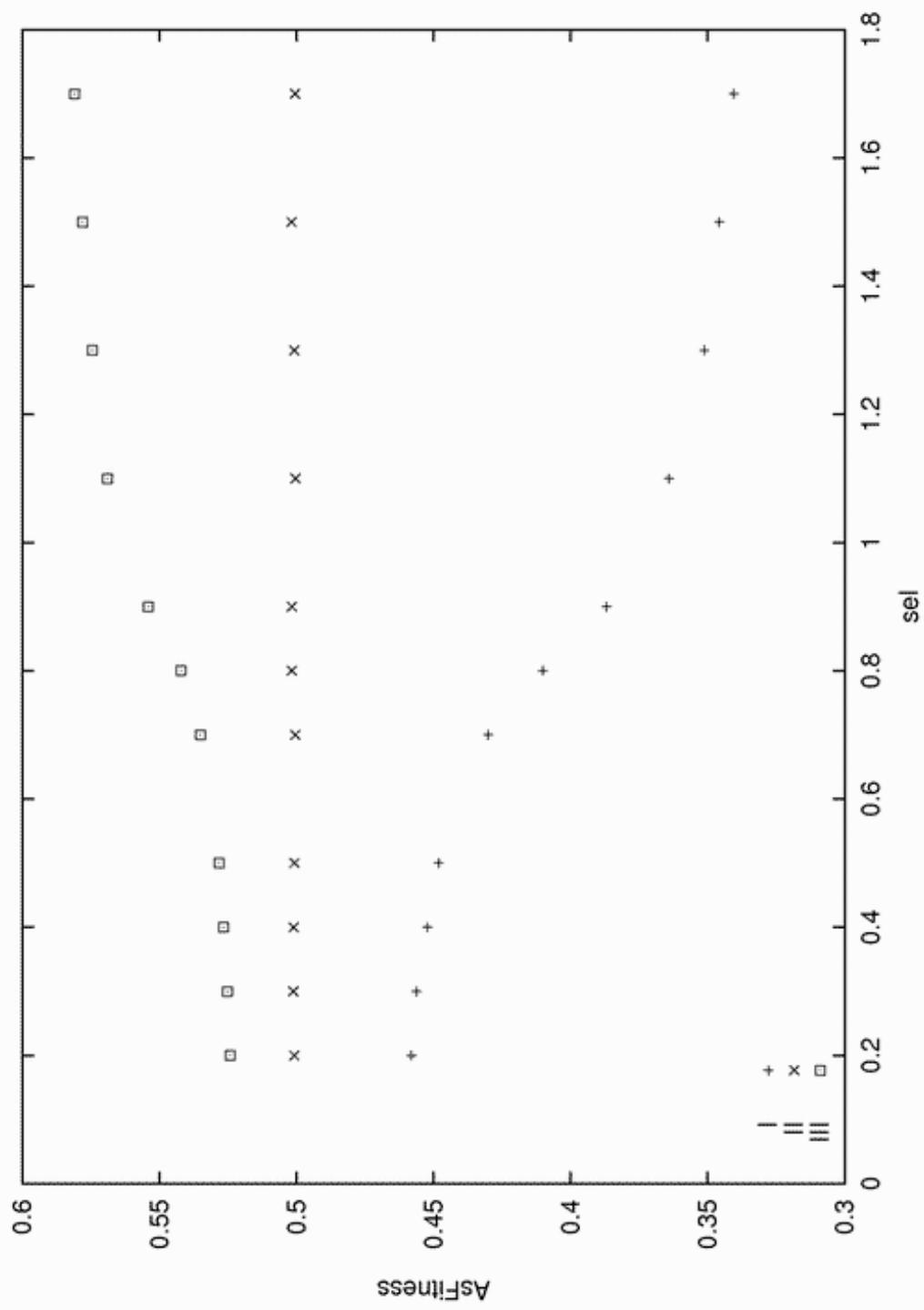
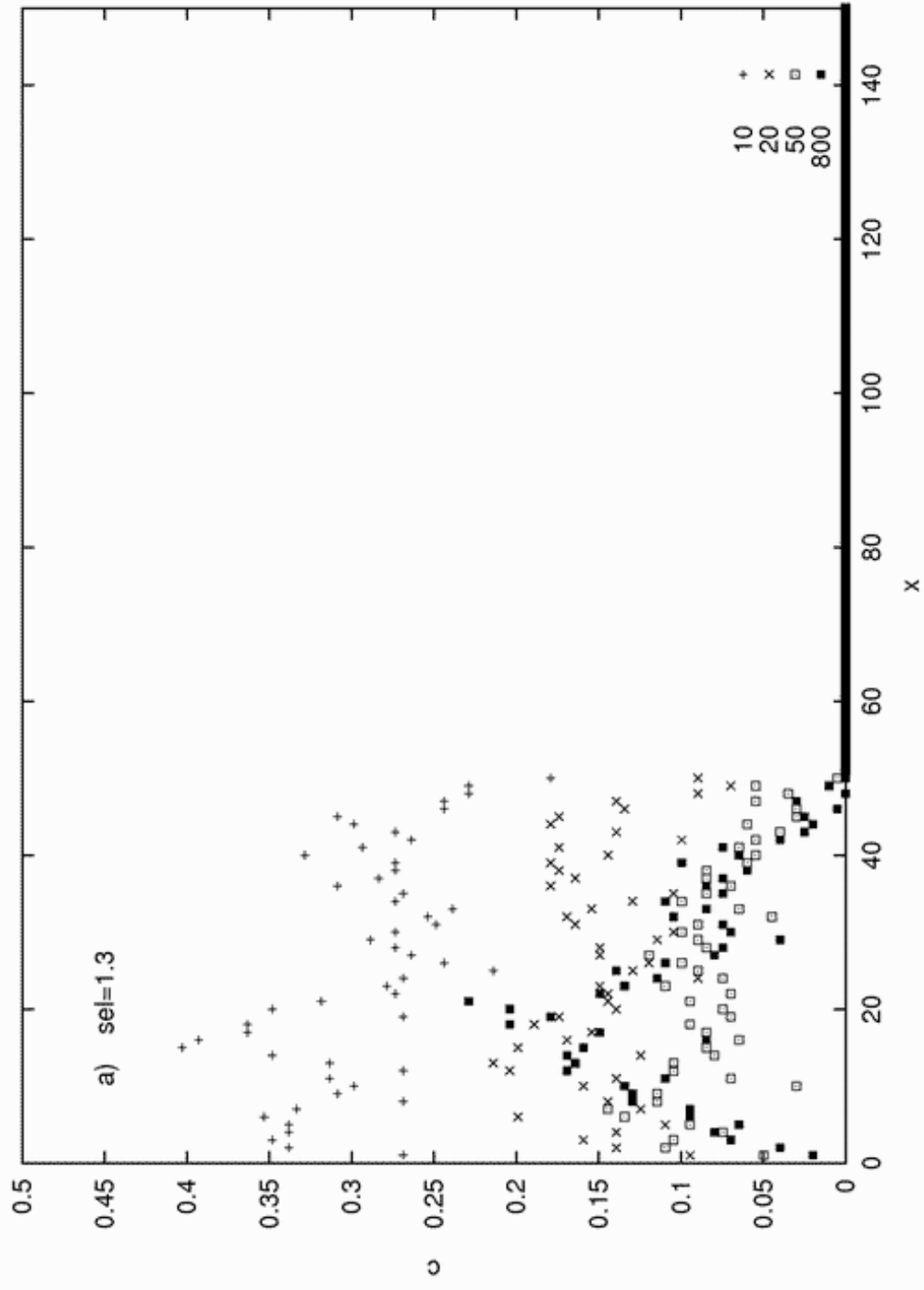
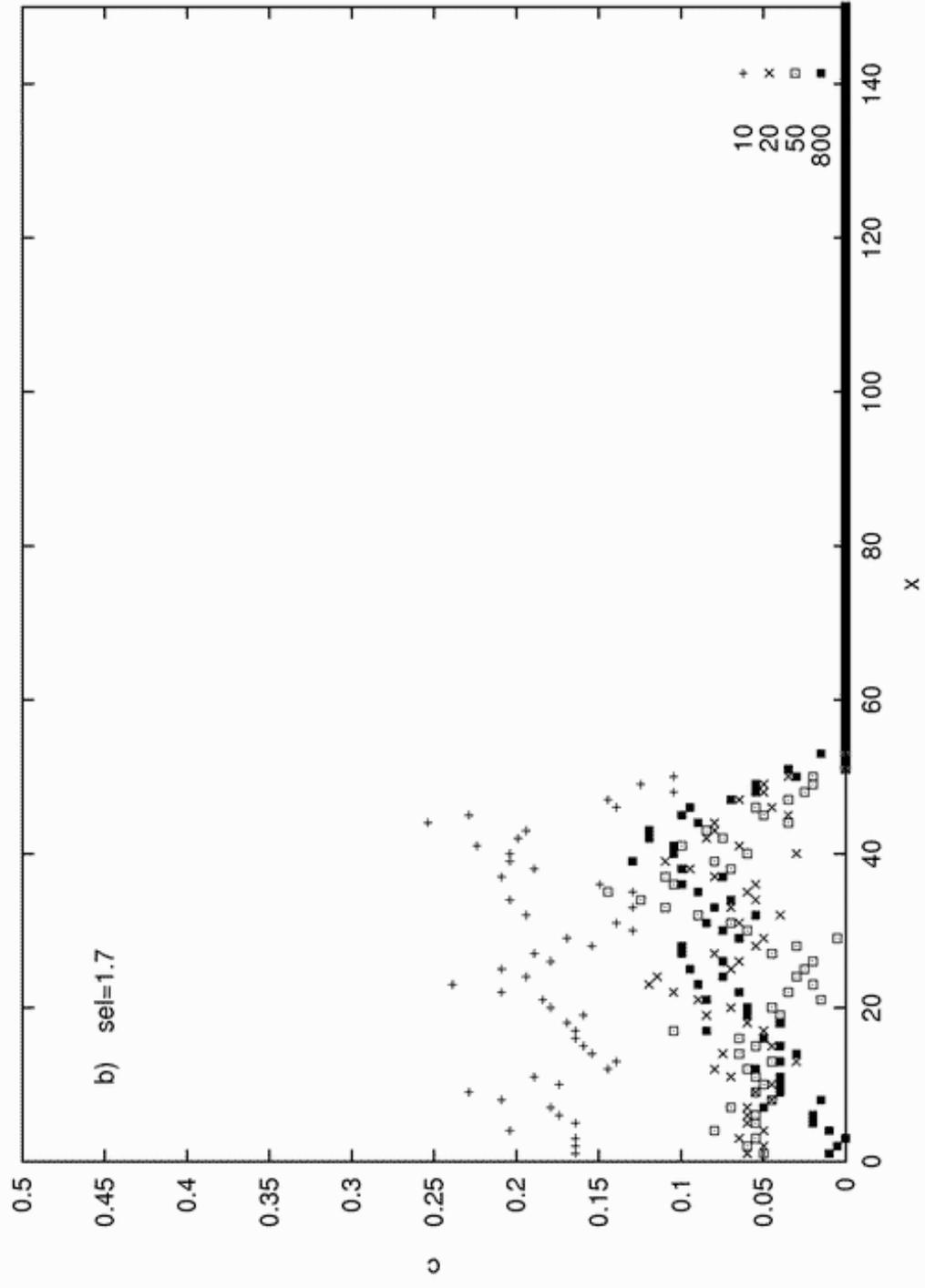
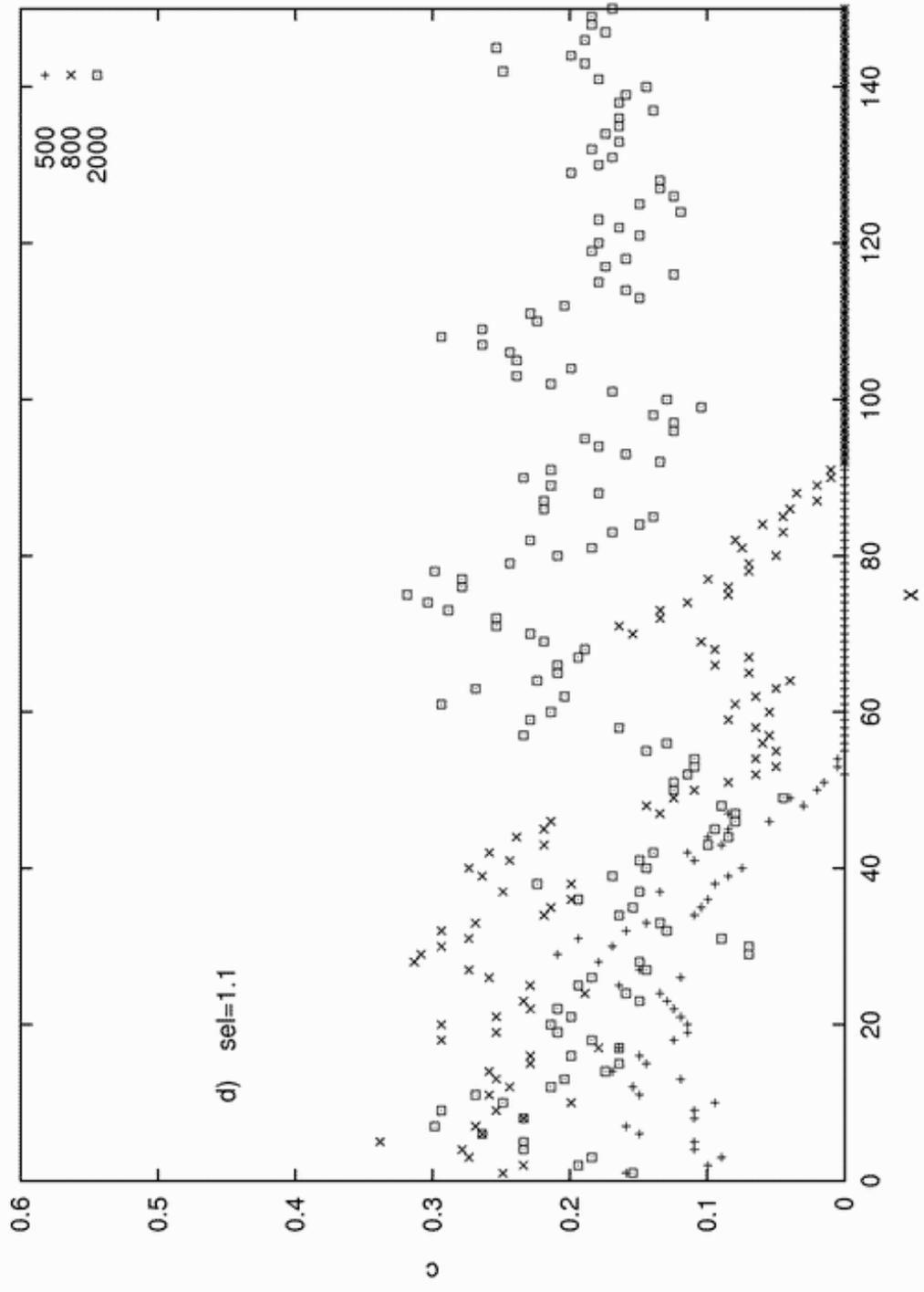
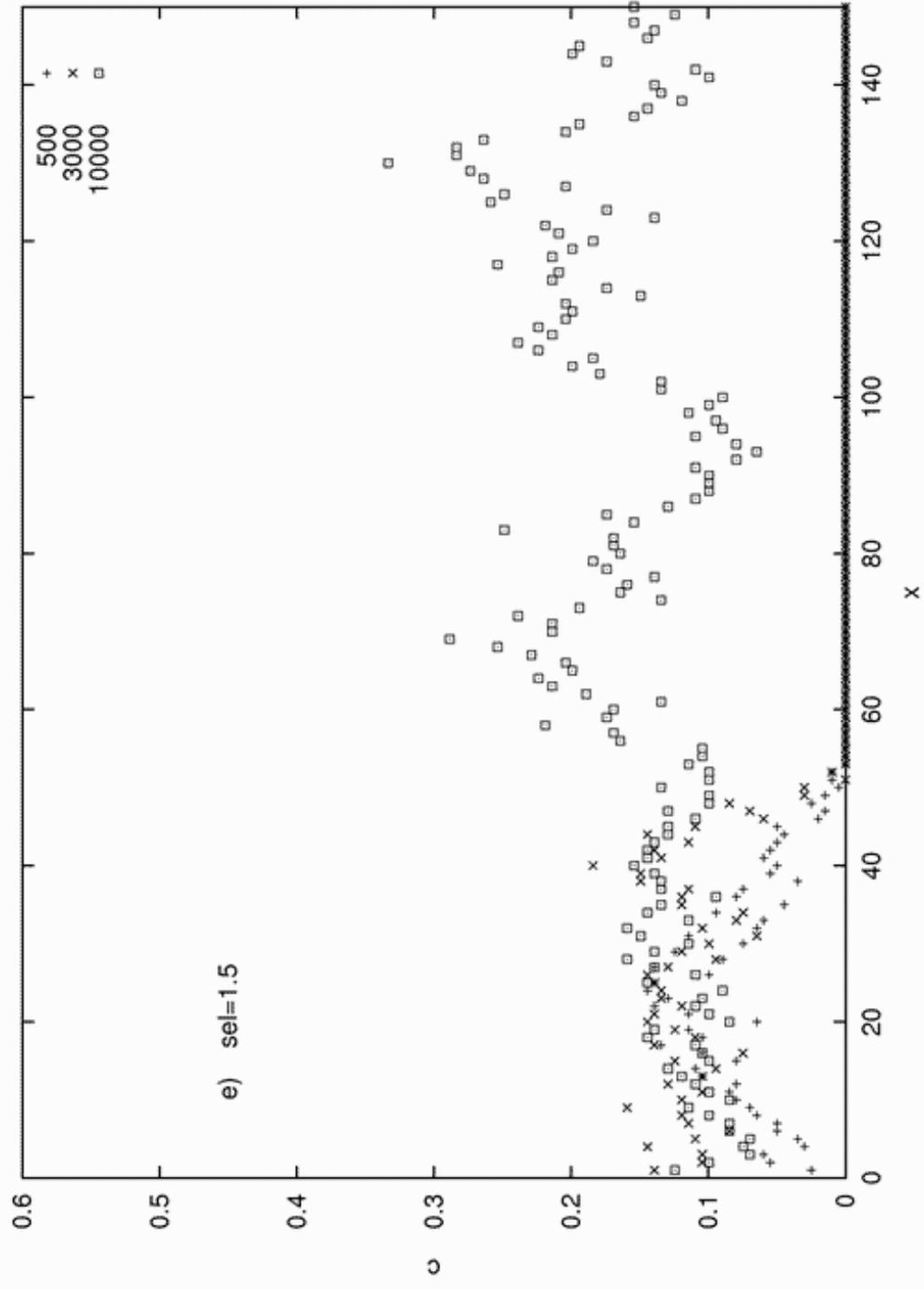
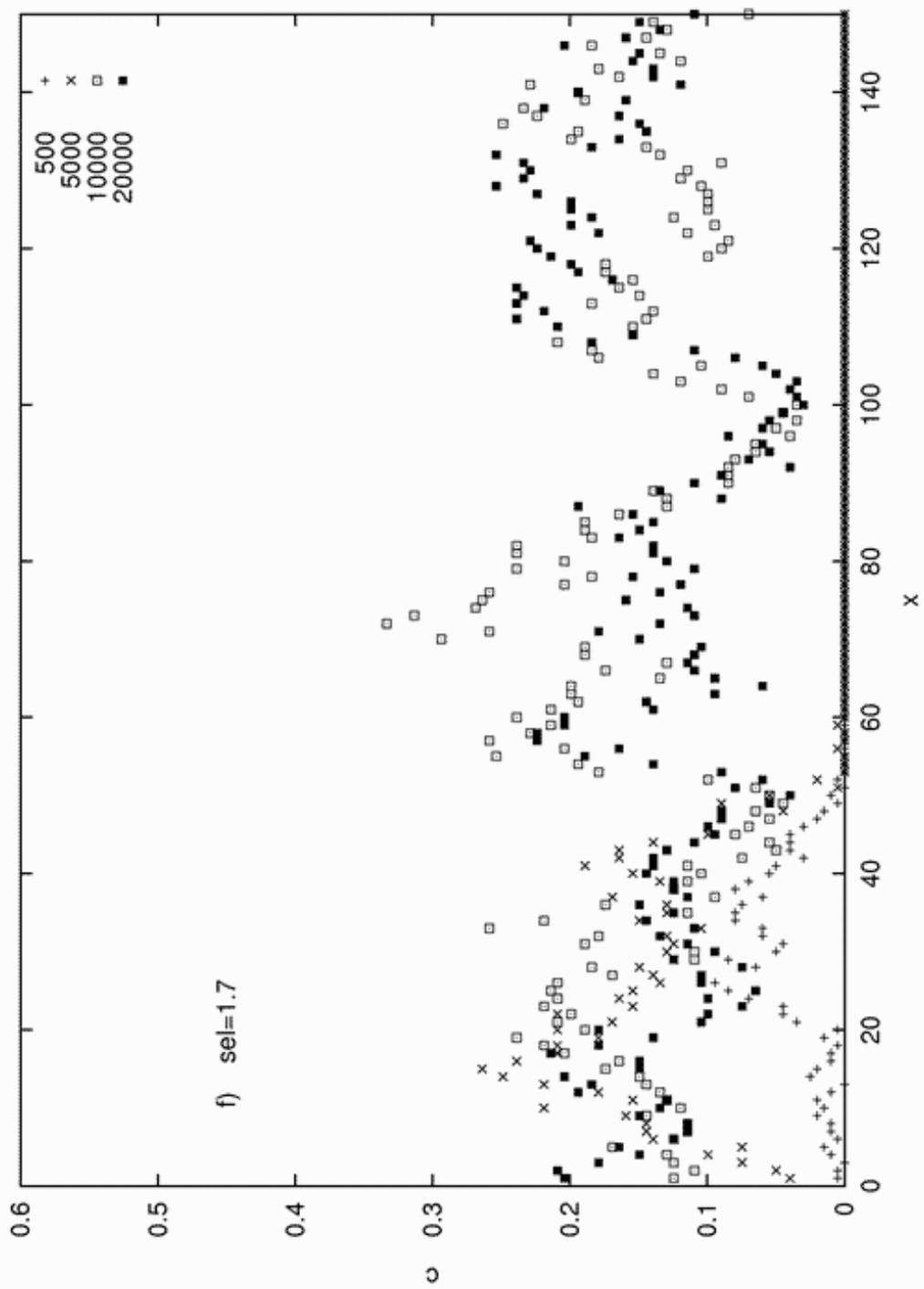
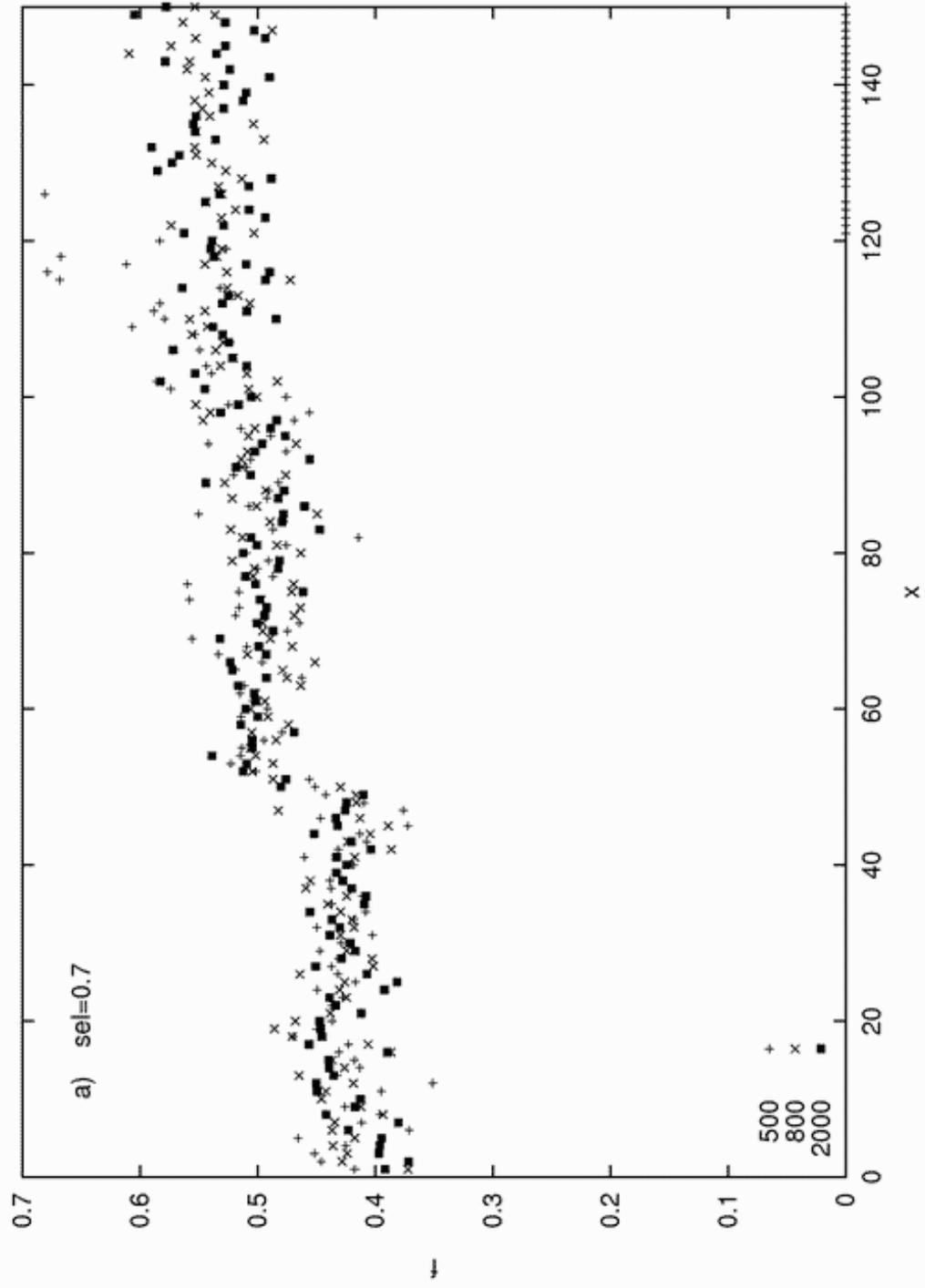
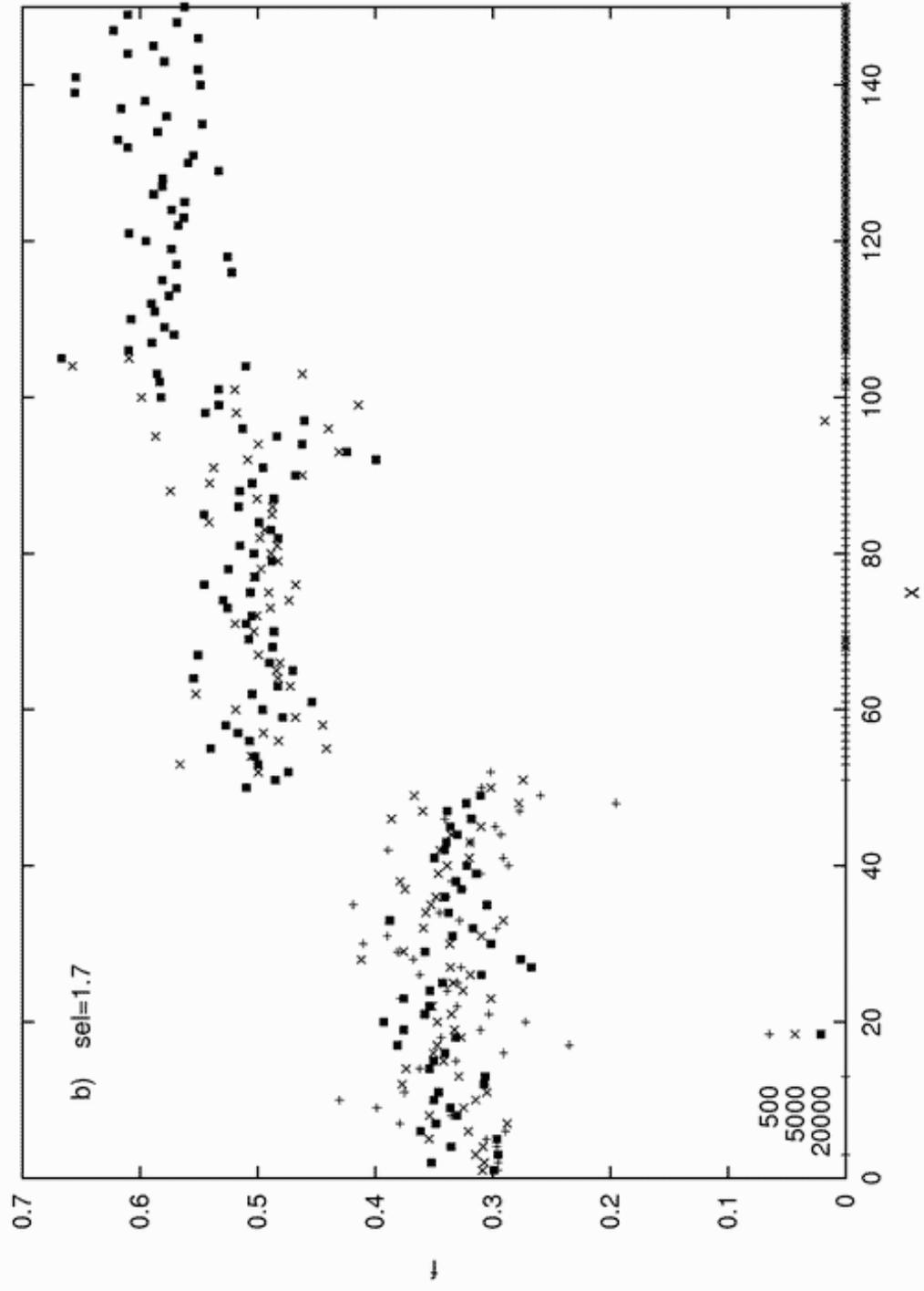
\end{document}